\documentclass{article}

\usepackage{arxiv}

\usepackage[utf8]{inputenc} 
\usepackage[T1]{fontenc}    
\usepackage{hyperref}       
\usepackage{url}            
\usepackage{booktabs}       
\usepackage{amsfonts}       
\usepackage{nicefrac}       
\usepackage{microtype}      
\usepackage{cleveref}       
\usepackage{lipsum}         
\usepackage{graphicx}
\usepackage{natbib}
\usepackage{doi}

\title{A novel neural network-based approach to derive a geomagnetic baseline for robust characterization of geomagnetic indices at mid-latitude}

\newif\ifuniqueAffiliation

\ifuniqueAffiliation 
\author{ \href{https://orcid.org/0000-0003-0937-2655}{\includegraphics[scale=0.06]{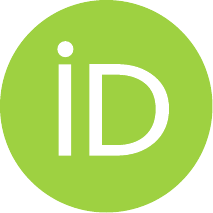}\hspace{1mm}Rungployphan Kieokaew}\\
	Institut de Recherche en Astrophysique et Plan{\'e}tologie,\\
	University of Toulouse 3, CNRS, CNES\\
	Toulouse, France \\
	\texttt{rkieokaew@irap.omp.eu} \\
	\And
	\href{https://orcid.org/0000-0002-9140-3627}{\includegraphics[scale=0.06]{orcid.pdf}\hspace{1mm}Veronika Haberle} \\
	Conrad Observatory, GeoSphere Austria,\\ 
	Vienna, Austria\\
	\And
	\href{https://orcid.org/0000-0002-1625-3462}{\includegraphics[scale=0.06]{orcid.pdf}\hspace{1mm}Aurélie Marchaudon} \\
	Institut de Recherche en Astrophysique et Plan{\'e}tologie,\\
	University of Toulouse 3, CNRS, CNES\\
	Toulouse, France \\
	\And
	\href{https://orcid.org/0000-0002-0458-8417}{\includegraphics[scale=0.06]{orcid.pdf}\hspace{1mm}Pierre-Louis Blelly} \\
	Institut de Recherche en Astrophysique et Plan{\'e}tologie,\\
	University of Toulouse 3, CNRS, CNES\\
	Toulouse, France \\	
	\And
	\href{https://orcid.org/0000-0001-8793-1315}{\includegraphics[scale=0.06]{orcid.pdf}\hspace{1mm}Aude Chambodut} \\
	Université de Strasbourg, CNRS, \\
	ITES UMR 7063, EOST UAR 830, \\
	Strasbourg, France
}
\else
\usepackage{authblk}

\newbox{\orcid}\sbox{\orcid}{\includegraphics[scale=0.06]{orcid.pdf}} 
\author[1]{%
	\href{https://orcid.org/0000-0003-0937-2655}{\usebox{\orcid}\hspace{1mm}Rungployphan Kieokaew\thanks{\texttt{rkieokaew@irap.omp.eu}}}%
}
\author[1,2]{%
	\href{https://orcid.org/0000-0002-9140-3627}{\usebox{\orcid}\hspace{1mm}Veronika Haberle}%
}
\author[1]{%
	\href{https://orcid.org/0000-0002-1625-3462}{\usebox{\orcid}\hspace{1mm}Aurélie Marchaudon}%
}
\author[1]{%
	\href{https://orcid.org/0000-0002-0458-8417}{\usebox{\orcid}\hspace{1mm}Pierre-Louis Blelly}%
}
\author[3,4]{%
	\href{https://orcid.org/0000-0001-8793-1315}{\usebox{\orcid}\hspace{1mm}Aude Chambodut}%
}
\affil[1]{Institut de Recherche en Astrophysique et Plan{\'e}tologie, UPS, CNRS, CNES, Toulouse, France.}
\affil[2]{Conrad Observatory, GeoSphere Austria, Vienna, Austria.}
\affil[3]{Université de Strasbourg, CNRS, ITES UMR 7063, F-67000 Strasbourg, France.}
\affil[4]{Université de Strasbourg, CNRS, EOST UAR 830, F-67000 Strasbourg, France.}
\fi

\renewcommand{\shorttitle}{Kieokaew, R., et al.}

\hypersetup{
pdftitle={A template for the arxiv style},
pdfsubject={q-bio.NC, q-bio.QM},
pdfauthor={David S.~Hippocampus, Elias D.~Striatum},
pdfkeywords={First keyword, Second keyword, More},
}

\begin{document}
\maketitle

\begin{abstract}
Geomagnetic indices derived from ground magnetic measurements characterize the intensity of solar-terrestrial interaction. The \textit{Kp} index derived from multiple magnetic observatories at mid-latitude has commonly been used for space weather operations. Yet, its temporal cadence is low and its intensity scale is crude. To derive a new generation of geomagnetic indices, it is desirable to establish a geomagnetic `baseline' that defines the quiet-level of activity without solar-driven perturbations. We present a new approach for deriving a baseline that represents the time-dependent quiet variations focusing on data from Chambon-la-Forêt, France. Using a filtering technique, the measurements are first decomposed into the above-diurnal variation and the sum of 24h, 12h, 8h, and 6h filters, called the daily variation. Using correlation tools and SHapley Additive exPlanations, we identify parameters that dominantly correlate with the daily variation. Here, we predict the daily `quiet' variation using a long short-term memory neural network trained using at least 11 years of data at 1h cadence. This predicted daily quiet variation is combined with linear extrapolation of the secular trend associated with the intrinsic geomagnetic variability, which dominates the above-diurnal variation, to yield a new geomagnetic baseline. Unlike the existing baselines, our baseline is insensitive to geomagnetic storms. It is thus suitable for defining geomagnetic indices that accurately reflect the intensity of solar-driven perturbations. Our methodology is quick to implement and scalable, making it suitable for real-time operation. Strategies for operational forecasting of our geomagnetic baseline 1 day and 27 days in advance are presented.
\end{abstract}


\section{Introduction}

Magnetic observatories at the ground level measure a superposition of magnetic fields of several sources at certain geographical locations. The dominant source is the Earth's intrinsic magnetic field, also called the ``main field", generated by geodynamo processes in the Earth's fluid inner core. The main field contributes over 93\% of the magnitude of the magnetic measurements at the surface, about tens of thousands of nano teslas (nT). Another internal source is the magnetized lithosphere which contributes with locally different but temporally nearly constant values \citep[e.g.,][]{2010SSRv..155...95T}. Other sources contributing to the geomagnetic field are electric currents flowing in the ionosphere and magnetosphere. In the ionosphere, the solar quiet (Sq) current in low- and mid-latitudes in the E-region is a dominant source that gives rise to the regular daily variations on the order of tens of nT \citep[e.g.,][]{2017SSRv..206..299Y}. It forms on the sunlit side as powered by the solar irradiance. The Sq variations are believed to be affected by tidal waves of atmospheric origins, which are global-scale oscillations with harmonic periods of a day. Along the magnetic equator, a strong zonal current forms a belt known as the equatorial electrojet \citep[EEJ;][]{1951AMGBA...4..368C}. At high latitudes, there are auroral electrojets \citep[AEJ;][]{https://doi.org/10.1029/JA080i025p03585, rostoker1979auroral} driven by the ionospheric-solar dynamo. Depending on the energy input by the solar wind through convection and particle precipitation, the auroral ionospheric conductivities vary and give rise to AEJ, marking the auroral ovals in the northern and southern hemispheres \citep[e.g.,][and references therein]{https://doi.org/10.1029/2020JA028905}. In the magnetosphere, current systems such as the ring current and field-aligned currents are significantly enhanced during solar events and modulate the geomagnetic field \citep[e.g.,][]{https://doi.org/10.1002/2017RG000590}.

The solar wind and the interplanetary magnetic field (IMF) interact with the Earth's magnetic field through complex couplings in several regions from the bow shock down to the ionosphere and the ground level. Ground magnetic measurements are thus valuable data sources for studying effects of the solar-driven disturbances on the magnetospheric and ionospheric systems. Solar-driven disturbances including solar storms affect the overall magnetospheric-ionospheric systems that enhance current systems and govern complex interaction among them. Solar storms are solar transient structures that can disturb the Earth's magnetic field temporarily and consequently trigger geomagnetic storms involving magnetic reconnection at the Earth's magnetopause and in the magnetotail. Interplanetary coronal mass ejection (ICME) is a major type of solar disturbances caused by an eruption on the solar surface. Earth-directed ICMEs have effects measurable on the ground from several hours up to a few days \citep[e.g.,][]{2016ApSS.361..253B, https://doi.org/10.1002/2017JA024006}. Corotating interaction region (CIR) is another transient structure formed when the fast solar wind originated from the Sun's coronal holes takes over a slower wind. The compression region and high-speed wind embedded in CIRs can also disturb the geomagnetic field up to several days \citep[e.g.,][]{https://doi.org/10.1002/2016JA023793, https://doi.org/10.1002/2016SW001559}. Characterization of the intensity or effects of these solar storms on the various systems is a vital task of the space weather community. 

Geomagnetic indices characterizing the intensity of solar-terrestrial activities are derived from ground magnetic measurements. \textit{K}-indices were first introduced by \citet{1939TeMAE..44..411B} to indicate the level of the perturbations with respect to a regular variation at a 3-hour range at mid-latitude. The \textit{K}-indices were derived for the Niemegk observatory with a scale of 0 (quiet) to 9 (strongly disturbed). The \textit{Kp} (\textit{K}-planetary) index was later derived from a network of 13 mid-latitude observatories \cite{bartels1949standardized}. Since their first conception, more geomagnetic indices have been proposed and concretized. Other \textit{K}-derived indices include \textit{aa} that was derived from two antipodal observatories from which the longest time series are available \citep[e.g.,][]{https://doi.org/10.1029/JA077i034p06870}. The \textit{am}, \textit{an}, and \textit{as} indices were proposed by \citet{Mayaud1968} to indicate sub-auroral magnetic activities at global, northern and southern scales. A comprehensive review of geomagnetic indices can be found in \citet{menvielle2010geomagnetic}. As an effort to improve the time resolution and scale of the \textit{Kp} index, the \textit{Hpo} indices were proposed at 30-minute and 1-hour resolutions \citep{2021SpWea..1902641M, https://doi.org/10.1029/2022GL098860}. Furthermore, new hemispheric geomagnetic indices $\alpha_{15}$ with 15-min resolution derived from a network of 48 mid-latitude observatories in the northern and southern hemispheres were proposed \citep{2015JGRA..120.9943C}.

To derive a new generation of geomagnetic indices, it is also desirable to establish a geomagnetic ``baseline" that characterizes quiet magnetic variations in the absence of solar disturbances. The quiet magnetic variations, defined as ``a smooth curve to be expected for that element on a magnetically quiet day, according to the season, the sunspot cycle and, in some cases, the phase of the Moon", traditionally involved hand-scaling by well-trained observers \citep{1939TeMAE..44..411B}. With the rise of the digital age, algorithms to automatically generate \textit{K} indices were proposed \citep[see][]{1995GeoJI.123..866M}. These algorithms involved an estimation of the non-\textit{K} variations that are the quiet variations according to the so-called Bartels-Mayaud rules \cite{118}. The Finnish Meteorological Institute (FMI) method \citep{sucksdorff1991computer} was found to be the most suitable for the continuation of \textit{K}-indices series without any serious jump in the statistics when passing from analog to numerical determination at one magnetic observatory \citep{1995GeoJI.123..866M}. Four algorithms including the FMI method have  been endorsed by the IAGA  (International Association of Geomagnetism and Aeronomy). Due to the lack of ground truth and clear identification of quiet sources, the subtraction of these empirically-derived baselines from ground magnetic measurements may not reflect the real intensity of solar-driven perturbations. This can have serious impacts on space weather applications and warnings. In this work, we aim to derive the quiet magnetic variation for definition of a new generation of geomagnetic indices with high time resolution and fine scale. As one of the first steps, we focus on the automated derivation of geomagnetic baseline only; the definition of new geomagnetic indices is left for future work.

To distinguish the perturbations of solar origin in the signals from other sources, establishing the geomagnetic baseline that robustly represents quiet periods is thus imperative. In an effort to derive a new magnetic activity index with a higher time resolution, \citet{2022JGRA..12730407H} proposed to characterize the magnetic measurements during quiet periods by filtering the signals into the above-diurnal ($>$24 hr), diurnal (24 hr), and sub-diurnal variations to capture physical sources at specific time scales and combine them to determine the geomagnetic baseline. This approach works rather well during quiet periods. It is efficient; it does not need any a priori information, thus it is scalable and suitable for near-real time applications. However, in the presence of solar-driven disturbances, the perturbations are present in all of the filters that were supposed to represent the quiet variations. Consequently, the actual intensity of solar-driven perturbations can be underestimated. Moreover, they compared their results to the FMI method. It turns out that the baseline from the FMI method follows the geomagnetic-storm variation similar to the filtering approach. There is thus still a need to robustly establish a baseline that contains quiet variations with minimal influence of storm perturbations. 

Machine learning neural networks have increasingly been used for space-weather related applications including prediction of magnetic activity indices \citep[e.g.,][]{https://doi.org/10.1029/2000JA900142, 2002GeoRL..29.2181L, 2014EP&S...66...95U, 2018SpWea..16.1882G, 2021SpWea..1902748C}. In this work, we consider an application of neural networks for automated generation of a geomagnetic baseline that can be used regardless of the solar (active or quiet) conditions. Our goal is to automatically produce a baseline that is not influenced by geomagnetic storms while robustly accounting for the main internal sources and the Sq variation and its possible day-to-day variability. We limit our focus to mid-latitude. Since \citet{2022JGRA..12730407H} have already decomposed the ground magnetic measurements to several contributions, we take these data as a starting point with the aim to demonstrate the capability of machine-learning based approach. Specifically, we consider using neural networks for time-series prediction as they allow us to consider independent parameters associated with physical sources contributing to the magnetic measurements. 

The organization of our paper is as follows. We first describe data and their pre-processing in Section~\ref{sec:data}. In Section~\ref{sec:feature-selection}, we explore correlations between the solar conditions and the geomagnetic measurements. Next, we describe the neural network architecture and the training approach in Section~\ref{sec:neural-network}. We then show the neural network results in Section~\ref{sec:nn-results}. Finally, we discuss the production of our new geomagnetic baseline and address our forecast strategies in Section~\ref{sec:baseline-forecast}. The summary and perspectives are presented in Section~\ref{sec:conclusions}. 

\section{Data}\label{sec:data}

We describe datasets used in this work comprising mainly ground magnetic field measurements collected from an observatory at mid-latitude and then pre-processed in order to capture and distinguish sources of their variation (Section~\ref{sec:ground-mag}). Since the geomagnetic field measurements are influenced by solar variabilities, we include in-situ monitoring of the solar conditions (Section~\ref{sec:solar-data}). As the geomagnetic field measurements are influenced by local time and seasons, we include proxies for such changes called ``geometrical parameters" (Section~\ref{sec:solar-data}). All these variables are useful for exploring relationships involving geomagnetic field responses to the solar, seasonal, and daily variabilities in Section~\ref{sec:feature-selection}. 

\subsection{Ground magnetic field data} \label{sec:ground-mag}

We focus on data from the magnetic observatory Chambon-la-Forêt (CLF) located at mid-latitude (48.0250N, 2.2600E) in France, Europe. The data are available at Bureau Central de Magnétisme Terrestre data repository from 1936 onwards. The data are replicated and associated with worldwide magnetic observatory data at the International Real-time Magnetic Observatory Network (INTERMAGNET) for the period from 1991 onwards. From \citet{2022JGRA..12730407H}, the data were processed from 1991 to 2019. The measurements were made at 1-min cadence. The data are provided in a local cartesian coordinate system (NED: North, East, Down). The \textit{X}-axis corresponds to the geographic north, the \textit{Y}-axis corresponds to the geographic east, and the \textit{Z}-axis completes the orthogonal system such that it directs towards the Earth's core.

In an effort to distinguish contributions from several sources to the ground magnetic measurements, \citet{2022JGRA..12730407H} first applied signal processing techniques to filter the measurement data. Using Finite Impulse Response filters, they decomposed the measurement data into the contributions at various time-scales. Firstly, the above-diurnal contribution correspond to the variation in the signals above 24 ($f_{>24}$) hours. Secondly, the diurnal and semi-diurnal contributions correspond to the variation at 24 ($f_{24}$) and 12 ($f_{12}$) hours, respectively. Finally, the contributions at 8 ($f_{8}$) and 6 ($f_{6}$) hours were also derived. To keep the same notation as \citet{2022JGRA..12730407H}, we call these various contributions as ``filter data". Using measurement data from observatories at low to mid latitudes in both hemispheres, they demonstrated that the derived filter data capture the physical sources contributing to the measurements reasonably well. The diurnal and semi-diurnal trends, in particular, are modulated by the season, the local time, and the day-to-day variation \citep[see][and references therein]{1989PApGe.131..315C}. \citet{2022JGRA..12730407H} combine all these filter data to determine a geomagnetic baseline during quiet periods. In this work, we use the sum of the diurnal harmonics: $f_D = f_{24} + f_{12} + f_{8} + f_{6} $; this newly-defined $f_D$, called ``daily filter", comprises the day-to-day, seasonal, and solar cycle variations. For the $f_{>24}$ variation, we extract the secular trend using linear regression. This will be detailed in Section~\ref{sec:extrapolation}.

Since using the full resolution (1-min) data in the neural networks is computationally expensive, as a first step, we consider using the filter data at a lower time cadence. Taking the original 1-min filter data, we perform a decimation. In essence, the decimation consists of obtaining the data at every hour, i.e., at every HH:00 where HH is a given hour from 01, 02, 03, ... to 23. To produce the $(f_D)$ at 1-hour cadence, we first sum $f_{24}$, $f_{12}$, $f_8$, and $f_6$ at the original 1-min cadence before decimating them. The decimation is chosen instead of an averaging to avoid over-processing of the data as the latter can remove useful signals.

\subsection{Solar wind and solar radio flux data} \label{sec:solar-data}

Solar wind conditions and solar variabilities drive the perturbation in the geomagnetic field. To get parameters relevant to these conditions, we utilize data products from the in-situ observations made upstream of the Earth at the Lagrangian L1 point as follows. We obtain the solar wind magnetic field and plasma datasets that are time-shifted to the Earth's bow shock nose \citep{2005JGRA..110.2104K} from CDAWeb (Coordinated Data Analysis Web). Specifically, we use 1-hour merged OMNI data product, available from January 1, 1995. The IMF data were obtained in the geocentric solar magnetic (GSM) coordinates, labelled as $B_x$, $B_y$, and $B_z$, where \textit{X}-axis points towards the Sun, \textit{Z}-axis corresponds to the geomagnetic north, and \textit{Y}-axis completes the right-hand orthonormal system. The plasma parameters were obtained for the proton bulk flow speed ($V$), the proton number density ($N$), and the proton temperature (\textit{Temp}). Also, we obtain the daily 10.7 cm solar radio flux ($F_{10.7}$) from the OMNI combined, definitive, and hourly product. The $F_{10.7}$ is an important indicator of the solar activity, derived from a measurement of the flux density computed from the total emission at 10.7 cm wavelength from all sources present on the solar disk made over 1 hour period \citep{2013SpWea..11..394T}.

\subsection{Geometrical data} \label{subsec:geom-param} \label{sec:geom-data}

Measurements at a magnetic observatory are influenced by the geographical location of the station (i.e. northern/southern hemisphere), the local time (i.e. day/night), and the season (i.e. the position of Earth around the Sun). Thus, parameters that record these variabilities, so-called ``geometrical parameters" are relevant. We chose the solar zenith angle (\textit{SZA}), the solar longitude (\textit{Ls}), and the distance between the Sun and the Earth (\textit{DistSE}). In addition, we derive the hourly local time (\textit{LT}), from 0 to 23, from the time stamps of the data. The \textit{SZA} is the angle measured from directly above the observation point (zenith) to the elevation of the Sun in the sky, measured from the horizon and is dependent on the geographical position of the station. The \textit{Ls} is the ecliptic longitude of the Sun; it indicates the position of the Earth around the Sun which relates to the seasons. The \textit{Ls} is defined as $0^o$ at spring equinox in the northern hemisphere, $90^o$ at summer solstice, $180^o$ at autumn equinox, and $270^o$ at winter solstice. The \textit{DistSE} is given in astronomical units (AU). All these parameters are indicative of daily and seasonal variations.

\section{Feature selection for the daily quiet variation} \label{sec:feature-selection}

To select relevant parameters for neural network modeling of the daily quiet variation, we first explore relationships between independent and dependent variables. The independent variables include the solar wind and IMF conditions, $F_{10.7}$, and the geometrical parameters. The dependent variables are the geomagnetic field variations; here we consider only the daily filter $f_D = (X_D, Y_D, Z_D)$ contribution. We note that the independent variables explored here are indicative of physical processes that drive the magnetic perturbations, but they are not necessary the ``drivers" of the perturbations. For instance, the geometrical parameters are not the drivers of the daily and seasonal variations. Indeed, they can be considered as ``proxies" of physical processes as they correlate to the Sq variation measured by a ground magnetometer \citep[e.g.,][]{2017SSRv..206..299Y, 2022JGRA..12730407H}. In addition, due to our choices of parameters, we cannot explore relationships associated with other existing drivers such as atmospheric waves, which can also contribute to the day-to-day variation \citep{https://doi.org/10.1002/jgra.50265, gi-2-289-2013}. Nevertheless, we expect that our choices of parameters, chosen here because of their accessibility and availability, can be used to model the majority of the daily quiet variation which is believed to be driven by the Sq current.

The relationships between variables can be linear or nonlinear, or non-existent. Using simple correlation analyses, we first explore if there exist linear or nonlinear correlations among the variables at the time scale of 1 solar cycle to have a global overview in Section~\ref{sec:correlations}. We then exploit a machine learning-based method to explore correlations between parameters during an interval with adjacent quiet and perturbed periods in Section~\ref{sec:SHAP}.

\subsection{Linear correlation and mutual information} \label{sec:correlations}

We first explore whether there exist linear correlations between the independent and dependent variables. Given two variables $\mathbf{x} = [x_1, x_2, x_3, ..., x_n]$ and $\mathbf{y} = [y_1, y_2, y_3, ..., y_n]$, the linear correlation coefficients can be found from
\begin{equation}
r = \frac{\sum (x_i - \bar{x})(y_i - \bar{y})}{\sqrt{\sum (x_i - \bar{x})^2 \sum (y_i - \bar{y})^2 }}, \label{eq:pcc}
\end{equation}
where $r$ is the correlation coefficient, $i = 1, 2, 3, ..., n$ are the $i$th element of $\mathbf{x}$ and $\mathbf{y}$, and $\bar{x}$ and $\bar{y}$ are the mean values of $\mathbf{x}$ and $\mathbf{y}$, respectively. The perfect, positive linear correlation is found if $r = 1$, while the perfect, negative linear correlation is found if $r = -1$. There is no linear correlation if $r = 0$.

\begin{figure}[p]
\centering
\includegraphics[width=0.8\textwidth]{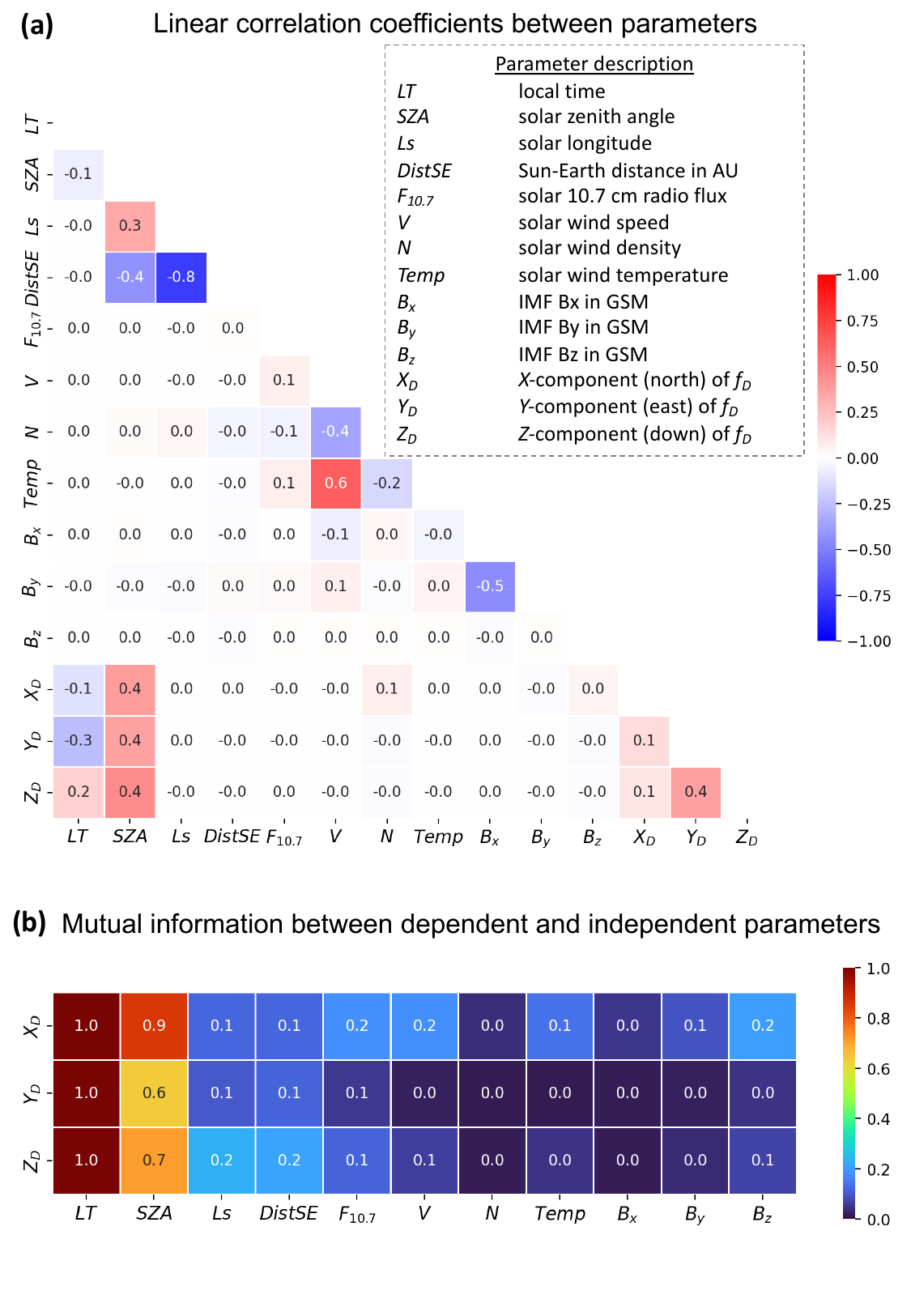}
\caption{(a) Linear correlation coefficients between parameters and (b) mutual information between the dependent parameters (vertical) and independent parameters (horizontal), using data between 1997 and 2007 (11 years in total).}
\label{fig:corr}
\end{figure}

We compute the linear correlation coefficients between the independent and dependent variables for the data between 1997 and 2007. Figure~\ref{fig:corr}a shows the linear correlation coefficients between the parameters highlighted with shaded colors: red and blue denote positive and negative correlations, respectively. We find that there are positive linear correlations between \textit{SZA} and $f_D = (X_D, Y_D, Z_D)$. \textit{LT} shows a weak, positive linear correlation with $Z_D$ while showing weak, negative correlations with $X_D$ and $Y_D$. This shows that \textit{LT} correlates with the various components of $f_D$ differently. There presents almost no, or very weak, correlation between $f_D$ and other independent parameters. Additionally, there are some correlations among the independent or dependent variables as they are confounding variables. For instance, \textit{DistSE} and \textit{Ls} both change with the season, and thus they show high linear correlation. Since both parameters contain the same information (seasonal variation), we may keep only one of them to remove redundancy. Here, we choose \textit{DistSE} because its variation is rather smooth unlike the \textit{Ls} which has a discontinuity when changed from $\sim359^o$ to $0^o$ at March equinox. The absence of clear linear correlation implies that they are either uncorrelated or they are correlated in the nonlinear way. Besides, as we consider a large amount of data, some linear correlations at shorter time scales, e.g., during solar maximum or minimum, may be hidden. For these reasons, we also consider more advanced approaches for exploring relationships between parameters next.

Although there is no linear correlation between certain parameters, there possibly is nonlinear correlation among them. The mutual information quantifies nonlinear dependency between two variables \citep{DIONISIO2004326}. Given two variables $\mathbf{x}$ and $\mathbf{y}$, the mutual information (\textit{MI}) is defined as
\begin{equation}
\textit{MI} = \sum_{j, k} p(x_j, y_k) \log \frac{p(x_j, y_k)}{p(x_j) p(y_k)}, \label{eq:mi}
\end{equation}
where $p(i)$ is the probability density function (PDF) of $i = \mathbf{x}, \mathbf{y}$ and $p(x, y)$ is the joint PDF of $\mathbf{x}$ and $\mathbf{y}$, and $j$ labels the $j$th element of $\mathbf{x}$ while $k$ labels the $k$th element of $\mathbf{y}$. The values of \textit{MI} are between $0$ and $1$ with $0$ being statistically independent sequences and $1$ being nonlinearly dependent sequences. Figure~\ref{fig:corr}b shows the \textit{MI} values with highlighted colors: dark red for $\textit{MI} = 1$ and dark blue for $\textit{MI} = 0$. We find that \textit{LT} and $f_D = (X_D, Y_D, Z_D)$ have perfect nonlinear dependencies. This means that $f_D$ has identical PDF to that of \textit{LT} (equation~\ref{eq:mi}). Also, \textit{SZA} shows strong nonlinear dependencies with the three components of $f_D$. Furthermore, we find that \textit{Ls,} \textit{DistSE}, and $F_{10.7}$ show some weak nonlinear dependencies with all components of $f_D$. Other solar wind and IMF parameters show weak or no nonlinear dependencies with $f_D$. We note again that the weak dependencies may be due to either the large time-scale considered or the absence of nonlinear relationships.

The analyses above allow us to explore relationships between the independent and dependent parameters. In brief, we find that all the three components of $f_D$ vary strongly with \textit{LT} and \textit{SZA}, both indicate the daily and seasonal variations. They also vary with $F_{10.7}$, \textit{DistSE}, and \textit{Ls} despite somewhat weak nonlinear correlations. The nonlinear dependency between $F_{10.7}$ and $f_D$ indicates the influence of the solar irradiance on the $f_D$ variation, especially for the $X_D$ component. Since \textit{DistSE} and \textit{Ls} provide similar information, i.e., the change in solar irradiance with respect to the Sun-Earth distance, we select only \textit{DistSE} to remove redundancy. This information is useful for selecting input parameters for the neural networks in Section~\ref{sec:neural-network}.

\subsection{SHapley Additive exPlanations (SHAP)} \label{sec:SHAP}

In addition to the correlation analyses, we may also explore relationships between independent and dependent variables based on certain \textit{ad hoc} diagnostic tools. The SHapley Additive exPlanations (SHAP) values \citep{NIPS2017_8a20a862} are among powerful tools that allow us to qualitatively explain predictions of machine learning (ML) models. This method is originally inspired by Shapley values in cooperative game theory \citep{doi:10.1073/pnas.39.10.1095} where coalitions of players form to achieve different profits in a cooperative game; thus, the technique can be used for fair allocation of credit to players in the game. With SHAP values, we can evaluate a ML model given a subset of input features and quantify each variable's or feature's impact in the model. The detailed explanation of SHAP values can be found in \citet{molnar2023interpreting}.

To quickly build a ML model for evaluating feature's impact using SHAP values, we consider using a surrogate model instead of the neural network itself. A surrogate model has the advantage to be quick to build; it is computationally light for training; and its optimization for reasonable results is relatively simple. Nevertheless, such a surrogate model may not be powerful or complex enough to provide robust results for time-dependent problems. Here we implement the XGBoost algorithm \citep{10.1145/2939672.2939785} that is a powerful tree boosting system for regression modeling. Since the $Y_D$ component typically shows the clearest daily and seasonal variations, we take only $Y_D$ as the target output. Using all the independent parameters as inputs and $Y_D$ as output, we train the XGBoost model using 11 years of the data prior to the interval of interest (see next). In the following, we discuss the XGBoost modeling results and the analyses using SHAP values.

\begin{figure}
\centering
\includegraphics[width=0.8\textwidth]{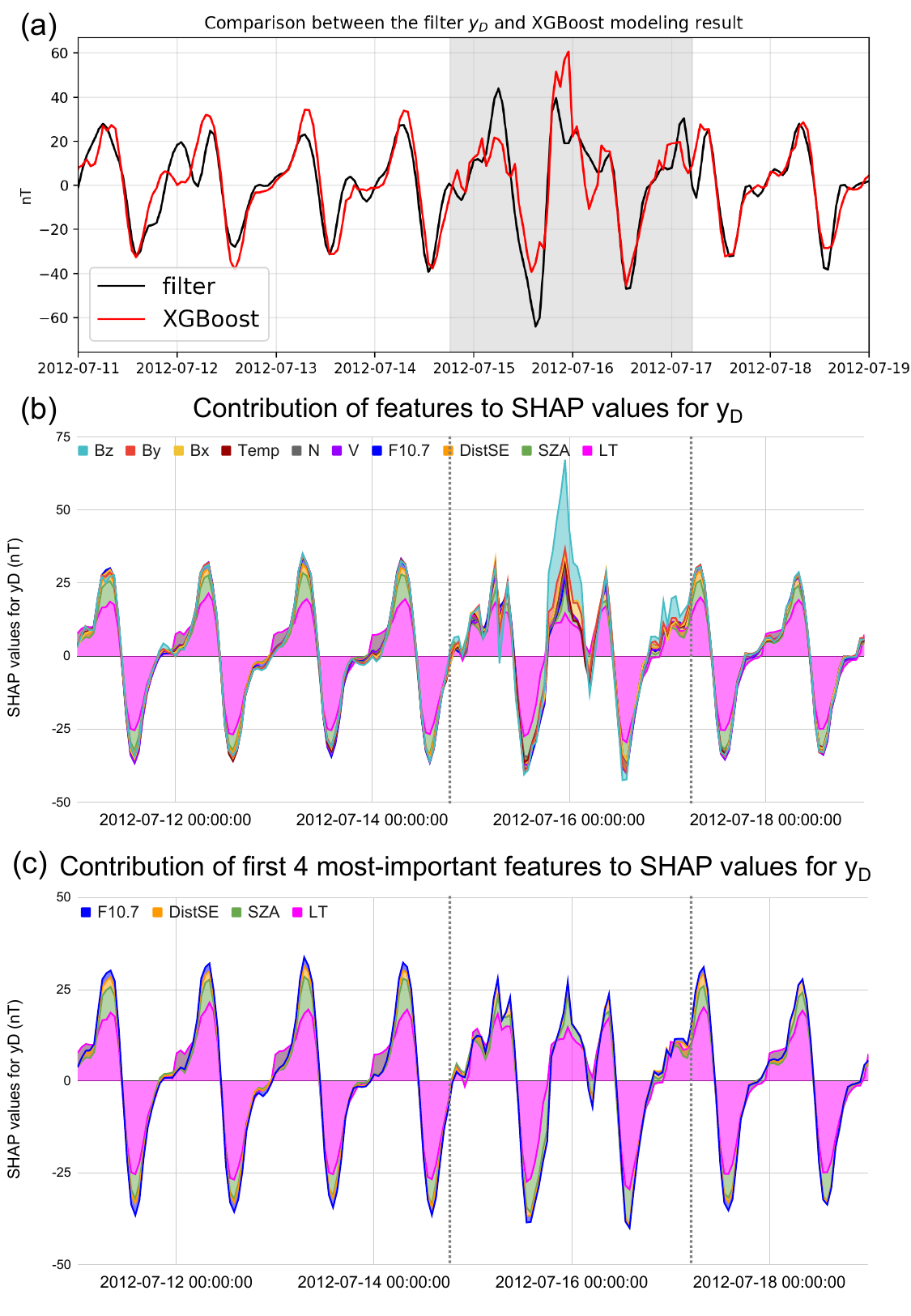}
\caption{(a) Comparison between $Y_D$ (black) and the modeled $Y_D$ (red) using XGBoost for the interval between July 11 and 19, 2012, including an ICME passage between July 14 at 18:09 and July 17 at 05:00 as shaded in grey. (b) Contribution of the input features to SHAP values for the modeled $Y_D$. (c) Contribution of the four most-important features to SHAP values for the modeled $Y_D$. The ICME passage is delineated by grey dotted lines in panels (b) and (c). }
\label{fig:XGBoost-shap}
\end{figure}

Figure~\ref{fig:XGBoost-shap}a shows a comparison between $Y_D$ (black) and the modeling result (red) using XGBoost. The data are from July 11 to 19, 2012, where the geomagnetic condition was mostly quiet, apart from July 14 at 18:09 to July 17 at 05:00 as shaded in grey where there was a passage of an ICME \citep{DVN/C2MHTH_2024}. The XGBoost model shows a good agreement with the $Y_D$ variation especially during the quiet period with the daily minimum and maximum being reproduced by the model. During the ICME passage, the daily minima and maxima of $Y_D$ appear enhanced. The XGBoost results, meanwhile, show somewhat poorer agreement at the extrema. Nevertheless, as the model produces globally correct results while being numerically light, we may take this model as a surrogate model for the exploitation of SHAP values.

We now evaluate contribution of the individual input variables through SHAP values to explain the modeling results of $Y_D$. Figure~\ref{fig:XGBoost-shap}b shows a stacked plot of the cumulative contribution of all the independent variables highlighted with different colors as a function of time. At each time instance, the cumulative contribution of all independent parameters is equal to the modeled $Y_D$ value in physical units (nT), which can be positive or negative. When considering several time instances, i.e., a time interval, the contribution of each independent variable is proportional to its (absolute) area under the graph; the larger the area under the curve, the more important the variable is. In general, we find that the most important contribution comes from \textit{LT} as shown in magenta. During the perturbed time delineated by the grey dotted lines, we find that the contribution of the input variables is different from the quiet time. Here we find that the contribution from the IMF Bz shown in cyan becomes significant during the ICME passage. This shows that the perturbed interval is driven differently. In the following, we consider the most important contributions for both quiet and perturbed times in order to exclude irregular contribution (e.g., driven by IMF) from the regular contribution.

Figure~\ref{fig:XGBoost-shap}c shows the first four most-important contribution to the SHAP values of $Y_D$ as modeled by XGBoost. This is done by ranking the input variables with the largest (absolute) area under the curve; then we limit the cumulative contribution to above $80 \%$ of the total area under the curve. In addition to the \textit{LT}'s contribution in magenta, the second most important contribution comes from the \textit{SZA} as highlighted in green. This shows that the modeled $Y_D$ varies the most with \textit{LT} and SZA. We note that, unlike \textit{LT}, \textit{SZA} contains also the seasonal variation in addition to the daily variation. The third most-important contribution appears to be \textit{DistSE} as shown in orange, which varies with the solar irradiance with the highest irradiance during northern-hemisphere summer (i.e., Sun-Earth perigee). Lastly, the fourth most-important contribution appears to come from $F_{10.7}$ as shown in blue, which varies with the solar activities following the solar cycle. In short, our analyses with SHAP help to identify important variables that most likely contribute to the modeled $Y_D$ using the surrogate ML model. The most important variables that contribute to the majority of the modeling results, regardless of the quiet or perturbed period, are found to be \textit{LT}, SZA, \textit{DistSE}, and $F_{10.7}$. We may use these four parameters to model the quiet, regular variation in Section~\ref{sec:neural-network}.

Our findings above are indeed consistent with previous findings on the Sq current system, believed to be the main contribution to the daily quiet variation, as measured by ground magnetometers at low- and mid-latitudes. Firstly, $F_{10.7}$ was found to be highly linearly correlated with the amplitude of Sq current \citep[e.g.,][]{2014JGRA..119.6732Y} as it is an excellent proxy of the solar activity indicative of the solar extreme ultraviolet (EUV) radiation, which influences the ionospheric conductivity. Secondly, the Sq current has seasonal dependency contained in \textit{SZA} and \textit{DistSE} in terms of its amplitude and position. During summer, the Sq amplitude is higher than during winter due to prolonged solar irradiation, which also influences the ionospheric conductivity \citep[e.g.,][]{1999JASTP..61..765T}. Finally, \textit{LT} was generally found to correlate with the Sq current measurement as the Sq cell is formed on the sunlit-side (daytime) \citep[e.g.,][]{2008AnGeo..26.1767S}. In brief, given that the daily quiet variation is mostly driven by the Sq current, these four parameters are reasonable choices for modeling the daily quiet variation that depends on the solar cycle, season, and daily variations.

\section{Neural network for the daily quiet variation} \label{sec:neural-network}

We now turn our attention to the prediction of $f_D$ without the solar transient perturbations using neural networks. Since there is no ground truth for the daily quiet variation, we employ $f_D$ during quiet time as a proxy for it. Indeed, \citet{2022JGRA..12730407H} demonstrated that $f_D$ can be used to produce a geomagnetic baseline in the absence of solar-driven perturbations as it comprises the Sq current and atmospheric contributions. In the presence of solar-driven perturbations, $f_D$ can no longer be used as a proxy for the ground truth, and we expect that our results would deviate from it. As shown in Section~\ref{sec:feature-selection}, we deduce that the most important parameters that correlate to $f_D$ regardless of the situations (while contributing the most during quiet conditions) are \textit{LT}, SZA, \textit{DistSE}, and $F_{10.7}$. Our choice for the neural network along with its brief functioning are described in Section~\ref{sec:nn-description}. We then outline its training in Section~\ref{subsec:model-training} before providing the results (Section~\ref{sec:nn-results}). 

\subsection{Neural network description} \label{sec:nn-description}

We develop a neural network with multiple input features and multiple output targets. The multiple input features are to accommodate the independent variables: LT, SZA, \textit{DistSE}, and $F_{10.7}$. The multiple output targets are set to accommodate the dependent variables consisting in the three components ($X_D$, $Y_D$, $Z_D$) of the daily filter. Since the ground magnetic measurements comprise the responses from the Sun and atmospheric conditions influencing the magnetospheric and ionospheric currents, the neural network must be able to account for the history of such conditions and/or physical processes. For this reason, we choose a recurrent neural network (RNN). Long Short-Term Memory networks (LSTM) are a variant of RNNs designed to overcome the vanishing gradient problems which typically arise with the long temporal dependencies \citep{10.1162/neco.1997.9.8.1735}. In principal, this type of neural network can keep track of the dependencies in the input sequences. Through the learning process, the neural network can memorize past input sequences that will likely affect the future data. Thanks to their efficiency, LSTM networks have now commonly been used in natural language processing and time series forecasting. Further description of the LSTM networks can be found in Appendix.

\begin{figure}[p]
\centering
\includegraphics[width=0.8\textwidth]{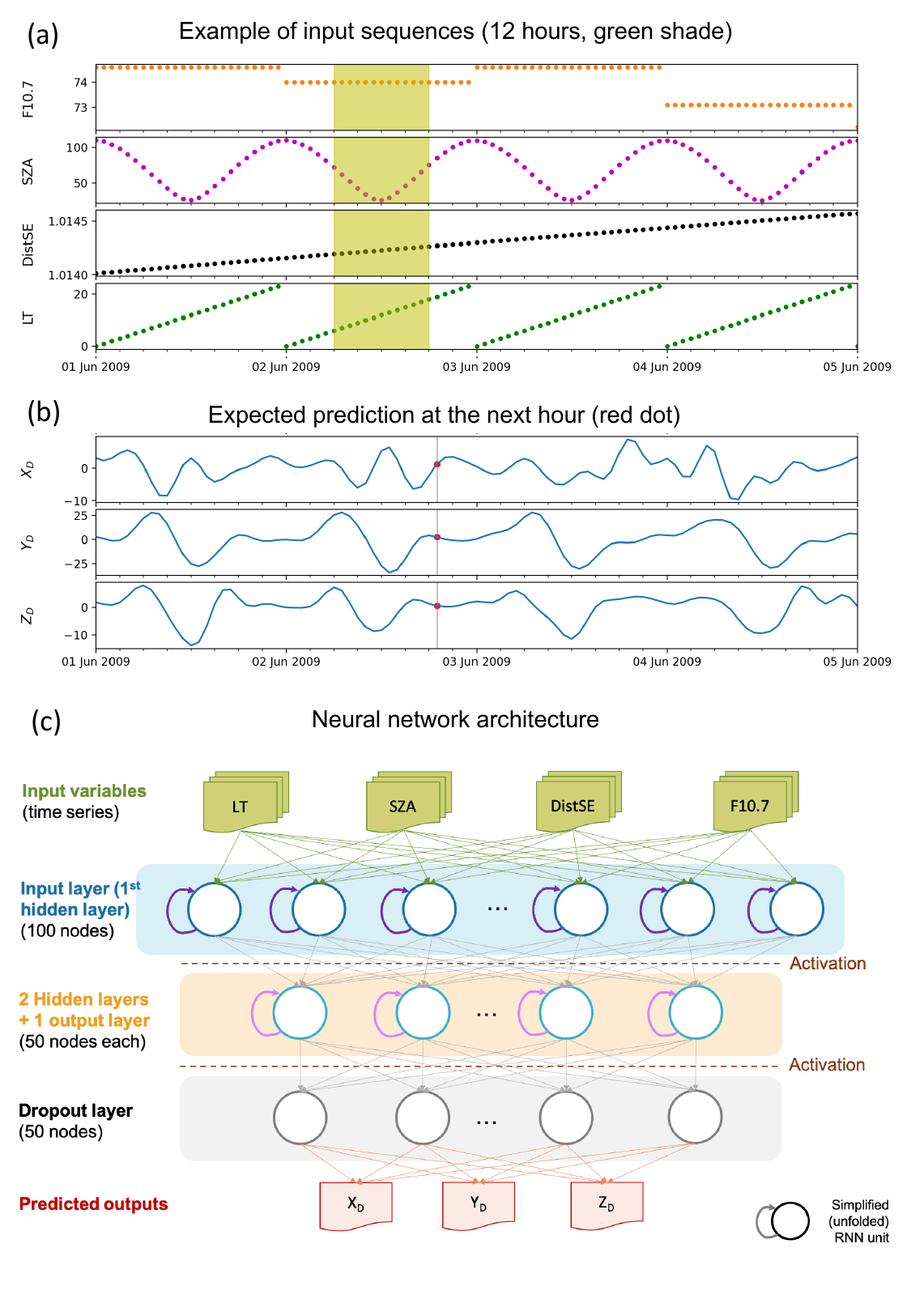}
\caption{(a) $F_{10.7}$, SZA, \textit{DistSE}, and \textit{LT} during June 1 and 5, 2009, highlighted in green as an example for the sequential inputs to the neural network. (b) $X_D$, $Y_D$, and $Z_D$ for the same interval, marked with red dots for the expected prediction. (c) Schematic of the neural network consisting of stacked layers and multiple RNN units (nodes), taken here as LSTM cells. The neural network takes in 12-hour sequences (green) of \textit{LT}, SZA, \textit{DistSE}, and $F_{10.7}$ up to the current time stamp $t_N$ and predicts the daily quiet variations (red) at the next adjacent hour $t_{N+1}$.}
\label{fig:LSTM-diagram}
\end{figure}

We now outline the concept for the daily quiet variation prediction. The purpose is to predict $f_D$ based on history of the input variables. Given the time $t_N$, where $N$ is a running index within a sequence, for instance, we want to predict $X_D$, $Y_D$, $Z_D$ at time $t_{N+1}$ using the past sequences of the input variables, which are available up to $t_N$. Figure~\ref{fig:LSTM-diagram}a shows an example of $F_{10.7}$, SZA, \textit{DistSE}, and \textit{LT} during June 1 and 5, 2009. In the green shade, we highlight, for instance, a time interval that would be taken as sequential inputs for the neural network. With these sequential inputs, we aim to predict $X_D$, $Y_D$, and $Z_D$ for the next hour adjacent to the highlighted interval as illustrated in Figure~\ref{fig:LSTM-diagram}b, marked by red dots. The prediction step can be repeated by advancing through the sequence from left to right in Figures~\ref{fig:LSTM-diagram}a and \ref{fig:LSTM-diagram}b to generate continuous prediction of $f_D$, as will be employed in Section~\ref{sec:nn-results}.

To build a neural network, we start with a simple architecture consisting of an input layer, a hidden layer, and an output layer. The input layer consisting of LSTM cells is to accommodate the sequential inputs of \textit{LT}, SZA, \textit{DistSE}, and $F_{10.7}$ with a certain length; the output layer is to predict $X_D$, $Y_D$, $Z_D$ at a next time step. We set up experiments using training data in 1991 - 2001 (11 years), validation data in 2002, and test data in 2003. The input and output data are rescaled to the range $[0,1]$ by finding the minimum and maximum values of each feature for the data in 1991 - 2001 using the formula:
\begin{equation}
x_{scaled} = \frac{x - x_{min}}{x_{max} - x_{min}},
\end{equation}
where $x$ represents values of the individual input and output variables. The scaling values (e.g., $x_{min}, x_{max}$) are saved for rescaling the output from the neural network in the final step. The same scaling is applied throughout this work. As for the architecture, first, we vary the number of the nodes in the input and output layers to find satisfactory results. Second, we add more hidden layers to improve the neural network performance. For each layer, a nonlinear ``activation function" is applied to the output to introduce nonlinearity. Here, the Rectified Linear Unit \citep{pmlr-v15-glorot11a} is used as the activation function as it overcomes the vanishing gradient problem in multiple-layer networks. Third, to prevent the neural network from overfitting, a dropout layer is introduced to drop units along with their connections during the training \citep{10.5555/2627435.2670313}. Several sets of different architectures with different numbers of hidden layers and neurons were tested (see Supplementary Information; SI). Besides, a set of varying time windows of the sequential inputs was also tested (see SI); the best results are found using 12 hours.

Figure~\ref{fig:LSTM-diagram}c summarizes the architecture that yields the minimum loss, taken as the mean-squared error (MSE). It consists of stacked layers of multiple LSTM nodes, represented as RNNs unit in the diagram. The input layer (i.e., the first hidden layer) has 100 nodes. There are three additional hidden layers (orange shade) including one LSTM output layer with 50 nodes each. Then, a dropout layer with 50 nodes (grey shade) is placed with a dropout ratio of $0.2$. In essence, this layer randomly drops $20\%$ of the nodes and connections from the previous layer to prevent overfitting before yielding the predicted outputs (see SI for tests with different dropout ratios). This neural network takes 26 minutes of CPU time to train for 11 years of data. We further optimize its training so that it is appropriate for our applications next.

\subsection{Neural network training}\label{subsec:model-training}

For the model training, we set up the neural network to learn in batches where it learns from a certain amount of data at a time. Here, the batch size is set to 256 for the training data at 1 hour cadence (95,945 data points for 11 years). The weights and biases in the neural network nodes and connections are initialized with random numbers and then updated through several cycles. The number of training cycles is known as ``epoch". The learning process is optimized and tracked through the loss function, which evaluates the model performance during each training epoch. Here, the loss function is set to be the MSE. The optimization during its learning, i.e., the iterative process for updating weights in the neural network, is performed using the stochastic gradient descent. Here, we employ the adaptive moment estimation \citep{Kingma2014AdamAM}, known as ``Adam" optimization algorithm, as it is rather computationally efficient for deep neural network training. The learning rate (the step size in the stochastic gradient descent) is set to be 0.001. We monitor the learning process through the validation loss. The learning process is stopped once there is no improvement in the validation loss for five consecutive epochs. The best model is saved when the validation loss reaches a minimum value before the training stops.

To effectively train the neural network model, we split the datasets as the following. Overall, we split the sequential data into the training, validation, and test sets. The validation set is used for evaluating and monitoring the model performance during the learning over several epochs. The ground magnetic measurements have temporal dependencies coming from the solar wind and solar dynamo (influencing the solar activities or phases). Therefore, the choice of training and validation data can introduce biases. Firstly, the neural network must be trained using a sufficient amount of data, in this case a complete solar cycle, so that it learns as much as possible. Secondly, since a best model is chosen based on the validation data, the choice of validation data can also introduce a bias. For example, if the model is validated and selected using an interval of data with active solar activities, i.e., during a solar maximum where the occurrence of ICMEs is high, the model may not be appropriate for use during the quiet solar activities, i.e., during a solar minimum where the occurrence of ICMEs is low. To minimize such a bias, we propose a new strategy for the model training as follows.

 \begin{figure}[ht]
 \centering
\includegraphics[width=0.8\textwidth]{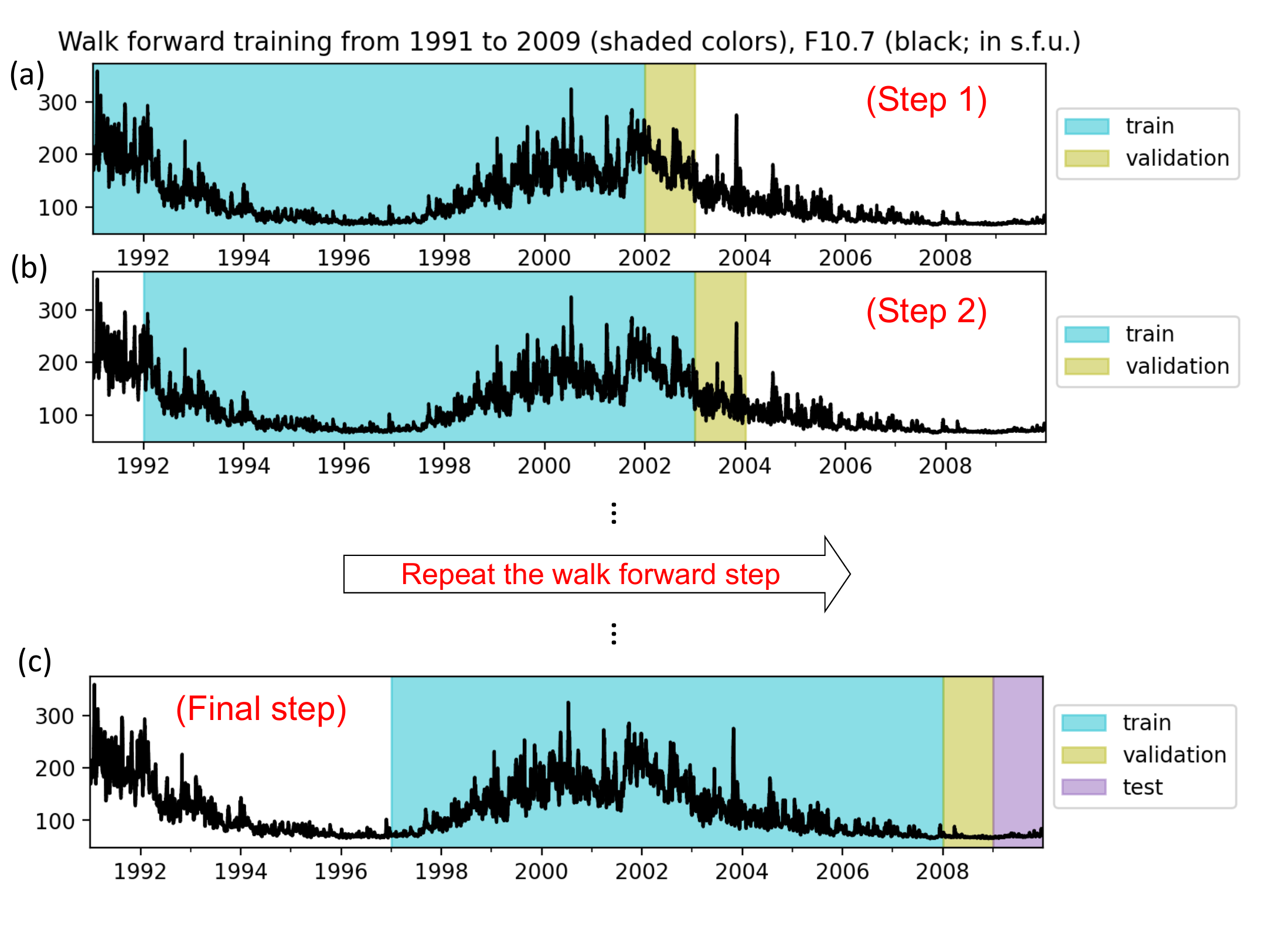}
 \caption{The walk forward training approach. The $F_{10.7}$ (black) indicative of the solar variability is shown for context. (a) Step 1 consists in training of the neural network with a specified training window (blue shade) and validating it with the adjacent data with a specified validation window (green shade). (b) Step 2 consists in updating the training with the next, shifted training and validation windows. The process is repeated up until to the test year. (c) The model is tested after the final training and validation step.}
 \label{fig:walkforward}
 \end{figure}

To best capture the different nature of solar activities in the various phases of the solar cycle, we propose an adaptive training method called ``Walk Forward Validation" (also called ``Sliding Window" or ``Rolling Forecast") approach \citep[e.g.,][]{brownlee2019introduction}. This approach has been used in economy and stock market predictions where the model is retrained once newer data become available \citep[e.g.,][]{KAASTRA1996215}. The advantage is that the model would be the most up-to-date, making it more relevant to the current situation and thus the near future situation. Essentially, the neural network is trained in several steps while moving forward along the time series. Figure~\ref{fig:walkforward} illustrates the walk forward training for the data starting from 1991 onwards. In each panel, we show $F_{10.7}$ indicative of the solar activity. We summarize our neural network training with the walk forward approach as follows.

\begin{itemize}
\item[1.] In Figure~\ref{fig:walkforward}a, the first step consists in the neural network training using the data within a specified minimum training window (11 years) as shaded in blue. It is then validated with the unseen data adjacent to the training data defined within a specified validation window (1 year) as shaded in green. 
\item[2.] In Figure~\ref{fig:walkforward}b, the trained neural network from step 1 is trained again using the data in a next, shifted 11-year window. The validation data in the previous step are included in the training data. The model is then validated with the unseen data, defined within a specified validation window, adjacent to the newly shifted training window. Technically, the weights and biases in the neural network are updated through this learning process using new data. 
\item[3.] The process is repeated until the end of all the training data excluding the test set. Figure~\ref{fig:walkforward}c shows the final training step with the test data shaded in purple. Here, for instance, the final model is validated with the data during low solar activities in 2008, therefore, it should be appropriate for the test using data in 2009 which also has low solar activities.
\end{itemize}

For our purpose, we define a minimum training window to be 11 years and a validation training window to be one year. Data in 2009 and 2012 are taken as the test datasets representative of the quiet and active solar periods, respectively. As shown in Figure~\ref{fig:walkforward}, the neural network was originally trained from 1991 to 2001, validated in 2002, and tested in 2003. Using the walk forward approach, we update the training up to 2009 with year 2008 being the validation data. Likewise for 2012, we update the training until 2012 with year 2011 being the validation data. Here, the model is most relevant to the time closer to the end of the training window as it is trained several times using the newer data, while being less relevant to the older data. This approach would offer optimum results for the time-dependent prediction. 

\section{Neural network prediction of the geomagnetic daily quiet variation} \label{sec:nn-results}

In this Section, we show prediction results for the daily quiet variation based on the neural network in Figure~\ref{fig:LSTM-diagram}, trained with the walk forward approach (see Figure~\ref{fig:walkforward}). On one hand, we expect our modeling results to be rather similar to the daily filter data during quiet days where the influence from solar-driven perturbations is minimal. On the other hand, we expect that our modeling results should be insensitive to solar-driven perturbations. These qualities are essential for characterizing a geomagnetic baseline, excluding the long-term secular variation, representative of the regular (daily) quiet variation in the absence of external drivers. In the following, we show first results during quiet period, i.e., during the solar minimum in 2009, and then results during non-quiet period, i.e., during the solar maximum in 2012.

As described in Section~\ref{sec:nn-description} (Figure~\ref{fig:LSTM-diagram}), the neural network prediction can be made one hour in advance only. For practical reasons, the results below are produced by running the neural network model consecutively over a certain period, e.g., for several days or weeks, to generate continuous prediction.

\subsection{Quiet period} \label{sec:quiet-time}

To characterize magnetic quietness with regard to the irregular geomagnetic activity, the International Service of Geomagnetic Indices (ISGI, \url{http://isgi.unistra.fr/}) provides the ``Really Quiet (C)" and ``Quiet (K)" days over 24 hours (CK24) or 48 hours (CK48). CK24 and CK48, deduced from $aa$ index \citep{https://doi.org/10.1029/JA077i034p06870}, are IAGA-endorsed data products. Here we focus only on CK48 days. The year 2009 has in fact the most CK48 days among the period between 1991 and 2020 (see Figure 2 of \citet{2022JGRA..12730407H}).

\begin{figure}[ht]
\centering
\includegraphics[width=0.8\textwidth]{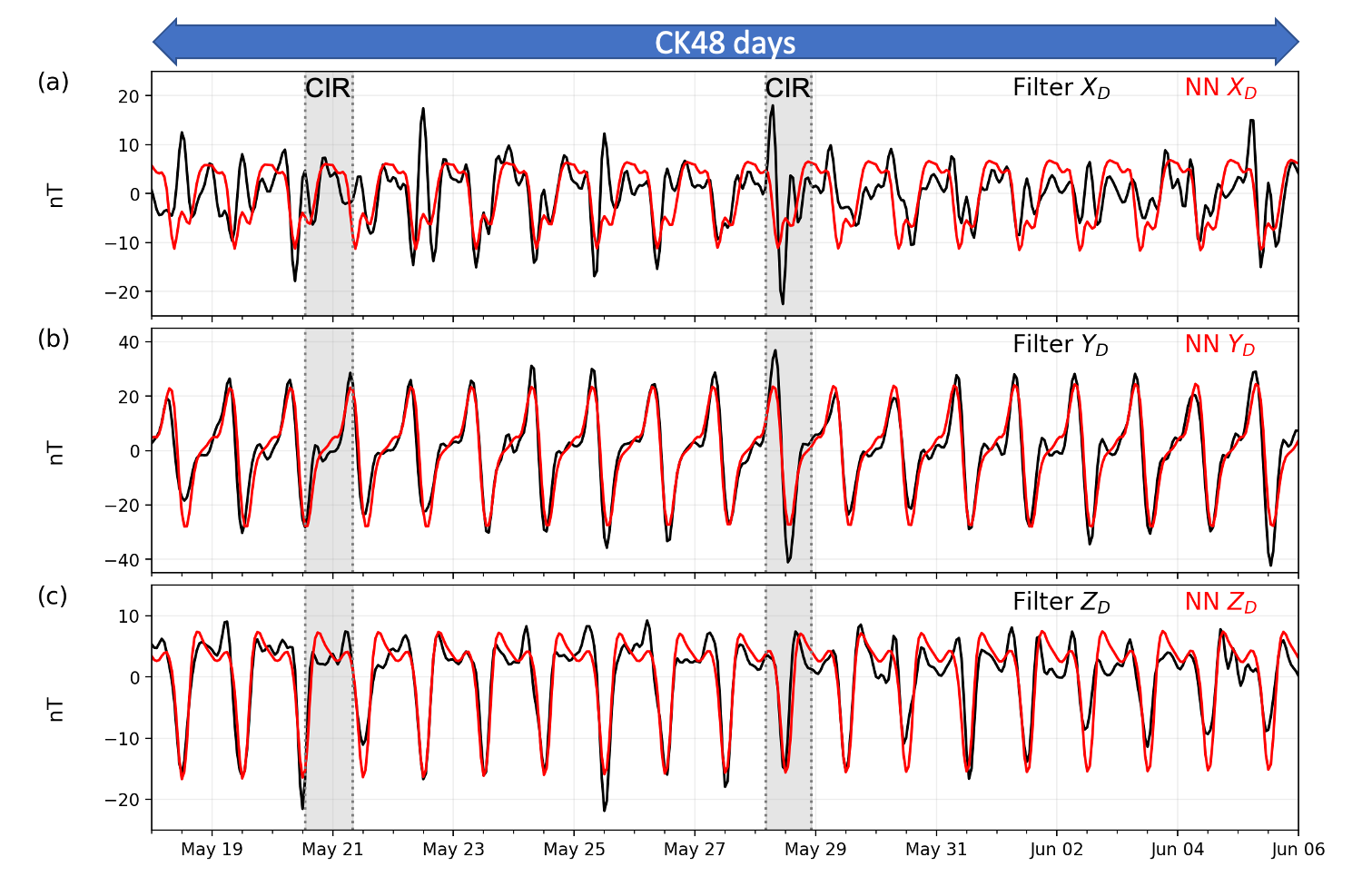}
 \caption{Comparison between the daily filter (black) and the neural network (NN) prediction of the daily quiet variation (red) during the consecutive CK48 days between May 18 and June 6, 2009, shown for (a) $X_D$, (b) $Y_D$, and (c) $Z_D$ components. Passages of CIRs on May 20 - 21 and May 28 are shaded in grey.}
 \label{fig:quiet-daily-baseline}
\end{figure}

We now focus on an interval with consecutive CK48 days from May 18 to June 6, 2009. Figure~\ref{fig:quiet-daily-baseline} shows a comparison between the neural network prediction for $X_D$, $Y_D$, and $Z_D$ (red) and the daily filter (black) in panels (a) - (c). During this period, there were two CIR arrivals on May 20 - 21 and May 28 \cite{Jian:2006ur} as shaded in grey. Overall, we find that the neural network results are qualitatively similar to the daily filter especially for $Y_D$. There are, however, some differences at small scales as well as for daily extrema. In fact, since the daily filter includes all signals at 24h, 12h, 8h, and 6h harmonics, all effects from atmospheric and tidal waves can be present. The absence of these features from the neural network prediction implies that there is no such information in the inputs or that these features cannot be learned. For $X_D$, the filter data show rather irregular variation although with some periodicity; our neural network shows rather different variation especially for the smaller-scale features. For $Z_D$, there are also some differences although less important than $X_D$.

On May 21, 22, and 28, the daily filter shows some enhanced extrema especially on the $X_D$ and $Y_D$ components, most-likely driven by the CIR passages. With CLF being at mid-latitude, we note that the magnetic measurements could be influenced by the perturbations coming from the higher latitudes such as AEJ as well as from the lower latitudes such as EEJ. These effects influence the $X_D$ and $Z_D$ components in particular, though their detailed mechanisms are not well understood. As the daily filter was derived from the signal decomposition, such perturbations can still persist in the data. For our neural network prediction, we find that the amplitudes of our daily quiet variation remain regular and do not vary with such solar-origin perturbations. This suggests that our neural network can be used to predict the daily quiet variation. Nevertheless, as discussed above, our neural network prediction may not include smaller-scale variations likely driven by atmospheric origins.

\subsection{Perturbed period} \label{sec:perturbed-time}

\begin{figure}[ht]
\centering
\includegraphics[width=0.8\textwidth]{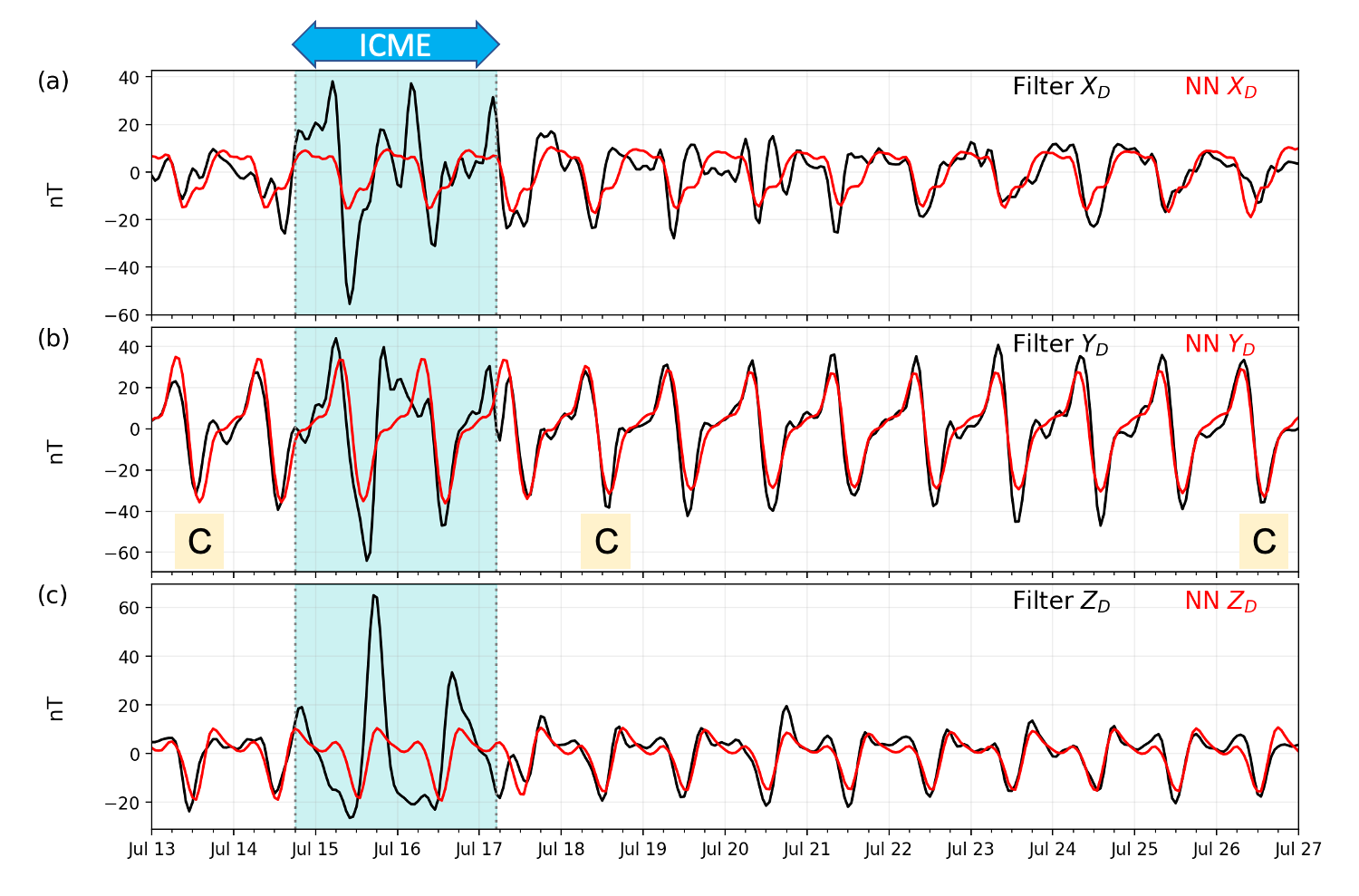}
 \caption{Comparison between the daily filter (black) and the neural network prediction (red) from July 13 to 27, 2012 for (a) $X_D$, (b) $Y_D$, and (c) $Z_D$ components. An ICME passage between July 14 at 18:09 and July 17 at 05:00, 2012, is shaded in blue. CK48 days are marked with ``C" in panel (b) on July 13, 18, and 26. } 
 \label{fig:quiet-daily-baseline-2012}
\end{figure}

We now turn our attention to the neural network prediction during non-quiet days. We select the interval from July 13 to 27, 2012, where a few CK48 days were present near a geomagnetic storm. Figure~\ref{fig:quiet-daily-baseline-2012} shows the daily filter (black) and the neural network prediction for the daily quiet variation (red) for $X_D$, $Y_D$, and $Z_D$ in panels (a) - (c). During this period, a geomagnetic storm was triggered following an Earth-directed ICME passage between July 14 at 18:09 and July 17 at 05:00, 2012, \citep{DVN/C2MHTH_2024} as shaded in blue. In addition, July 13, 18, and 26 are characterized by ISGI as CK48 days. They are marked by letter ``C" in Figure~\ref{fig:quiet-daily-baseline-2012}b.

During the ICME passage as shaded in blue, a clear perturbation is visible in the daily filter for all components, especially for $X_D$ and $Z_D$. Regarding the neural network prediction, our daily quiet variation shows regular variation that is insensitive to the geomagnetic storm for all components. Outside the storm time, in particular for CK48 days, our neural network prediction shows rather similar variation to the daily filter especially for $Y_D$ and $Z_D$. These demonstrate that our approach can be used to provide the daily quiet variation regardless of the presence of solar-origin perturbations unlike the filter method. Nevertheless, as discussed in Section~\ref{sec:quiet-time}, the neural network prediction does not include smaller-scale features (especially for $X_D$) and their extrema are not as enhanced as those of the daily filter. This may suggest again that our prediction does not include variation of atmospheric origins.

With the implementation in Section~\ref{sec:neural-network}, our method can yield the daily quiet variation 1 hour in advance. In Section~\ref{sec:baseline-forecast}, we consider extending the lead time of the forecast. In combination with the secular trend extrapolation, we also consider providing a new geomagnetic baseline forecast.


\section{Geomagnetic baseline forecasting} \label{sec:baseline-forecast}

To produce a geomagnetic baseline, we need two elements: (a) the daily quiet variation, and (b) the secular trend variation. For a geomagnetic baseline similar to the existing filter baseline or the FMI baseline, we need only $X$ and $Y$ components. In Section~\ref{sec:nn-results}, we demonstrated that the neural network can be used to provide the daily quiet variation 1 hour in advance. To provide a complete geomagnetic baseline, we also need to know the secular trend variation 1 hour in advance. We first consider this aspect in Section~\ref{sec:extrapolation}. Then, we propose a method to extend the forecast horizon to 1 day and 27 days in Section~\ref{sec:forecast-fmi}. Finally, we compare our proposed baseline with the filter baseline and FMI baseline.

\subsection{Extrapolation of the above-diurnal secular trend} \label{sec:extrapolation}

In addition to the daily quiet variation, we need to characterize the secular variation, i.e., excluding the solar-driven perturbation, in $X_{>24}$ and $Y_{>24}$. The secular trend variation is driven by the internal sources of the geomagnetic field including those originating in the core and lithosphere of the Earth \citep[e.g.,][]{2020SGeo...41.1611M}. The secular trend variation varies on the time scale between a month to a few thousand of years. On time scales of between a month and 100 years, the secular variation is entirely caused by the rigidly coupled movement of the magnetic field lines with the fluid motion in the liquid outer core (advection).

\begin{figure}[ht]
\centering
\includegraphics[width=0.8\textwidth]{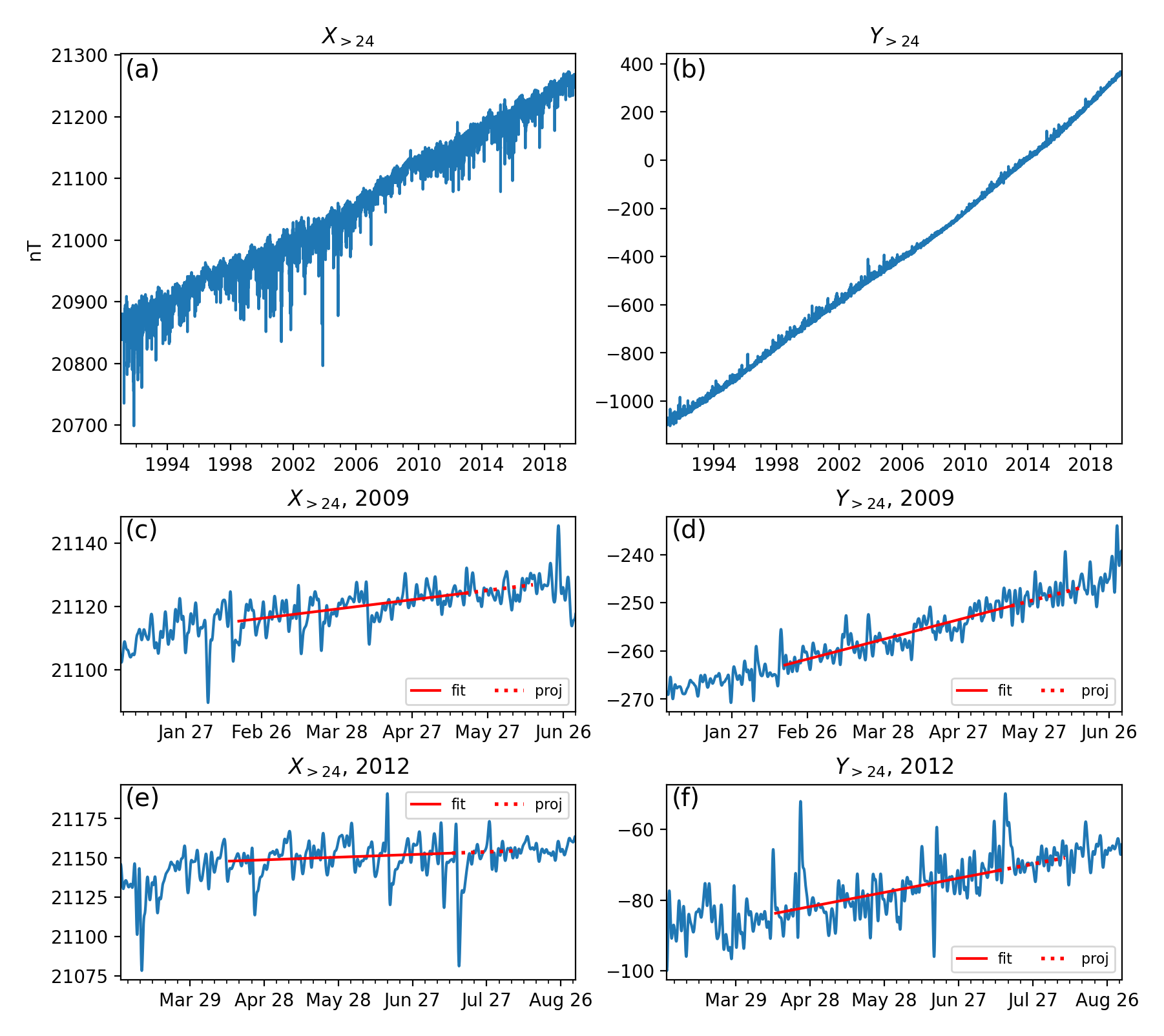}
 \caption{The above-diurnal filter derived from the measurements at CLF (blue). (a) $X_{>24}$ and (b) $Y_{>24}$ from 1991 to 2020. (c) - (f) Examples of linear fitting (red solid line) on $X_{>24}$ and $Y_{>24}$ and 27-day linear projection (red dashed line) for the chosen intervals in 2009 and 2012.}
 \label{fig:extrapolation}
\end{figure}

The secular variation has been recorded since mid 1500s at CLF \citep{2016HGSS....7...73M}. Figures~\ref{fig:extrapolation}a and \ref{fig:extrapolation}b show $X_{>24}$ and $Y_{>24}$ between 1991 and 2020 derived from the measurements at CLF. One can see that the secular trend appears rather linear to the first order. Over this time scale, $X_{>24}$ has increased for about 400 nT while $Y_{>24}$ has increased for about 1400 nT. On top of the secular variation, there are smaller-scale fluctuations owing to the solar-driven perturbations. For instance, one can see more-frequent drops in $X_{>24}$ in 1998 - 2006 during the high solar activity and less-frequent drops in $X_{>24}$ in 2008 - 2010 during the solar minimum. It would be desirable if we can characterize and project the secular trend based on physical understandings. For instance, we may use main geomagnetic field model outputs together with constants on each component, to take into consideration the local crustal biases at the considered magnetic observatory location. The IGRF model \citep[International Geomagnetic Reference Field;][]{2017JGRA..122.5687A}, updated every 5 years, in conjunction with constant values to consider the crustal field due to remnant rocks within the crust may be considered. The CHAOS-7 model of the geomagnetic field derived from observations from low-Earth orbit satellites, updated every 4 to 6 months, may also be used. Yet, these approaches will downgrade the capacity of real-time calculation and lead to use of a priori information, making it less convenient for operational implementation.

Assuming that the secular trend would remain linear in our time scale of consideration, e.g., for 1 hour up to 27 days, we may employ linear regression as follows. To obtain a projection for the next hour up to next 27 days, we perform linear fitting on $X_{>24}$ and $Y_{>24}$ individually, on the past data with a length of more than a month. Since the presence of solar-origin perturbations, typically lasting from a day up to a week, can deviate the trend, we take a longer interval in order to smooth such effects. We decide to take the data length of about three months or 90 days as the interval of $X_{>24}$ and $Y_{>24}$ to perform a linear fitting in order to project the secular trend. As an example, we perform linear fitting on the data preceding May 18, 2009, in Figures~\ref{fig:extrapolation}c and \ref{fig:extrapolation}d for $X_{>24}$ and $Y_{>24}$, respectively. The linear fitting on the past 90 days data is denoted by a red solid line. Next, we can project the same trend, using the same slope and intercept, up to 27 days as depicted by a red dashed line. Both the linear fitting and linear projection qualitatively agree with the global trend of the filter $X_{>24}$ and $Y_{>24}$. Similarly, we can perform the same linear fitting for the data preceding July 13, 2012, in Figures~\ref{fig:extrapolation}e and \ref{fig:extrapolation}f. These two examples demonstrate that the simple linear regression may be used to forecast the secular trend present in the above-diurnal filter locally.


\subsection{Forecast and comparison with FMI baseline} \label{sec:forecast-fmi}

Using the neural network for the daily quiet variation and the extrapolation method for the above-diurnal secular variation, we can combine them to obtain prediction of the geomagnetic baseline. From Section~\ref{subsec:model-training}, we obtain a neural network model that produces the daily quiet variation 1-hour in advance based on 12-hour history of LT, SZA, \textit{DistSE}, and $F_{10.7}$. In order to advance the lead time, we need to have $F_{10.7}$ forecast. Several approaches for forecasting $F_{10.7}$ exist \citep[e.g.,][]{https://doi.org/10.1029/2011SW000748, https://doi.org/10.1002/2017SW001637, refId0, https://doi.org/10.1155/2019/5604092, 10575844}. Nevertheless, to keep our approach ergonomic, we may simply use the standard $F_{10.7}$ data product with the following consideration. To obtain a forecast for the next day, we may assume that $F_{10.7}$ is similar to that of the current day. To provide a longer lead time of the forecast, the community often employs a recurrence model where the next solar rotation is assumed to have the solar irradiance similar to the past solar rotation \citep[e.g.,][]{frohlich2003solar, https://doi.org/10.1002/swe.20040}. With this consideration, we can use the $F_{10.7}$ over the past 27 days to provide a prediction over the next 27 days. 

\begin{figure}[ht]
\centerline{\includegraphics[width=0.8\textwidth]{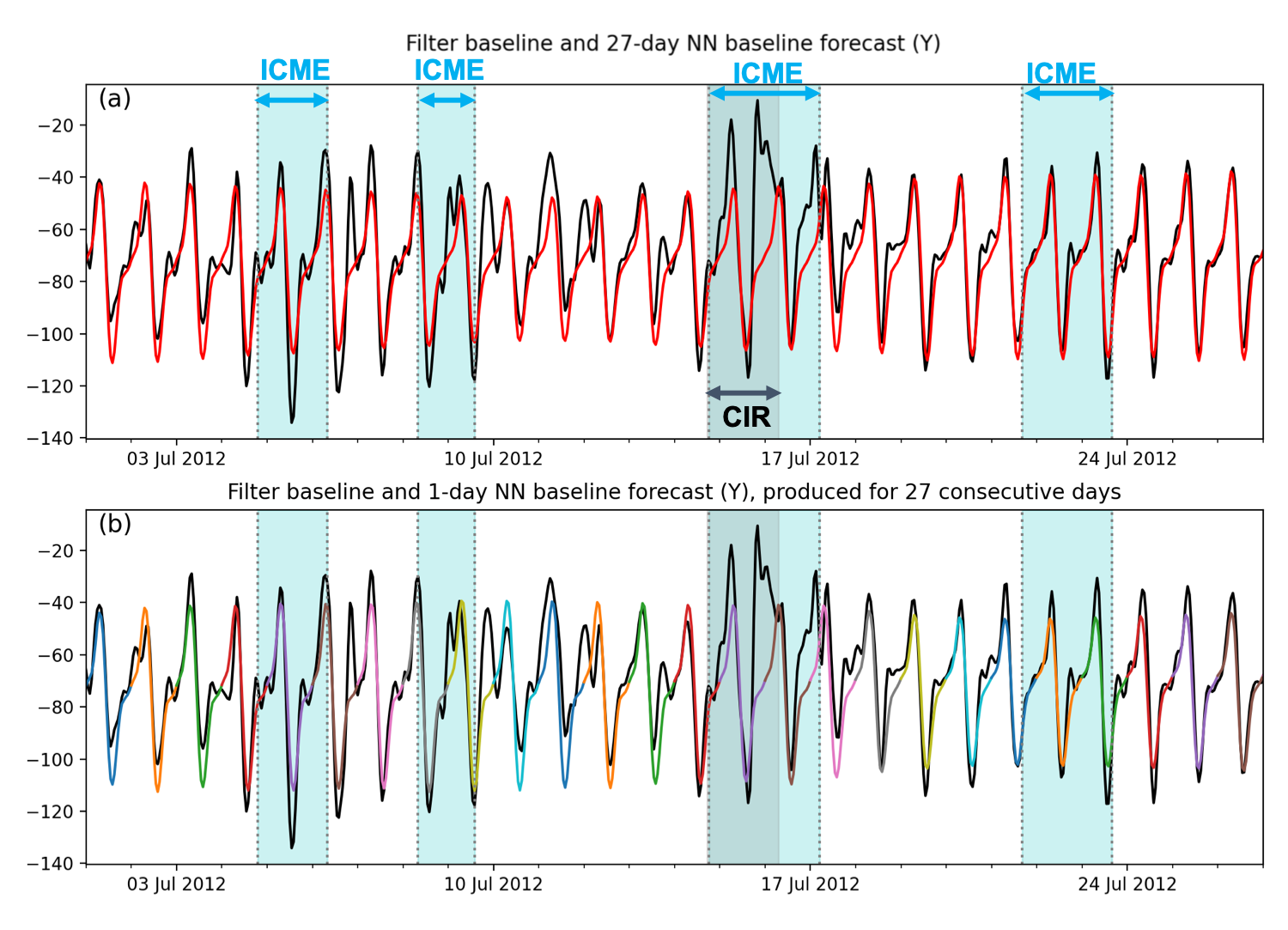}}
 \caption{Comparison between the filter baseline (black) and (a) the 27-day baseline forecast based on the neural network combined with the $f_{>24}$ trend linear projection (red), and (b) the 1-day baseline forecast produced daily for 27 consecutive days represented by different colors. Intervals during ICME passages are shaded in blue and an interval during a CIR passage is shaded in grey. } 
 \label{fig:forecast}
 \end{figure}

For the above-diurnal secular trend, we may also project the trend for the next hour, next day and next 27 days using the past 90 days of the data as shown in Section~\ref{sec:extrapolation}. In practice, all these procedures can be done daily to provide geomagnetic baseline forecast 1-day and 27-days in advance, in addition to the 1-hour baseline prediction using real-time $F_{10.7}$. Figure~\ref{fig:forecast}a shows an example of the 27-day forecast of the neural network (NN) produced baseline (red) combined with the secular trend projection for July 1 - 27, 2012. This is done by projecting the $F_{10.7}$ forecast to be similar to the past 27 days. During this period, there are four ICME passages \citep{DVN/C2MHTH_2024} as shaded in blue. Also, there is a CIR passage \citep{2022SoPh..297...30H} as shaded in grey embedded in the third ICME. Our baseline, labeled as `NN' (red), is shown in comparison to the filter baseline (black) for the $Y$-component. Similar to Section~\ref{sec:nn-results}, we find that our approach produces a variation that is insensitive to the solar-origin perturbations, seen as enhanced extrema of the filter baseline. In Figure~\ref{fig:forecast}b, we produce the 1-day baseline forecast that is updated daily at midnight for 27 consecutive days represented using different colors. This approach yields rather similar results as in Figure~\ref{fig:forecast}a (red) albeit some differences in the extrema. Therefore, our approach can be used to yield a new geomagnetic baseline forecast in advance for 1 day or 27 days, in addition to the real-time calculation.

\begin{figure}[ht]
\centerline{\includegraphics[width=0.8\textwidth]{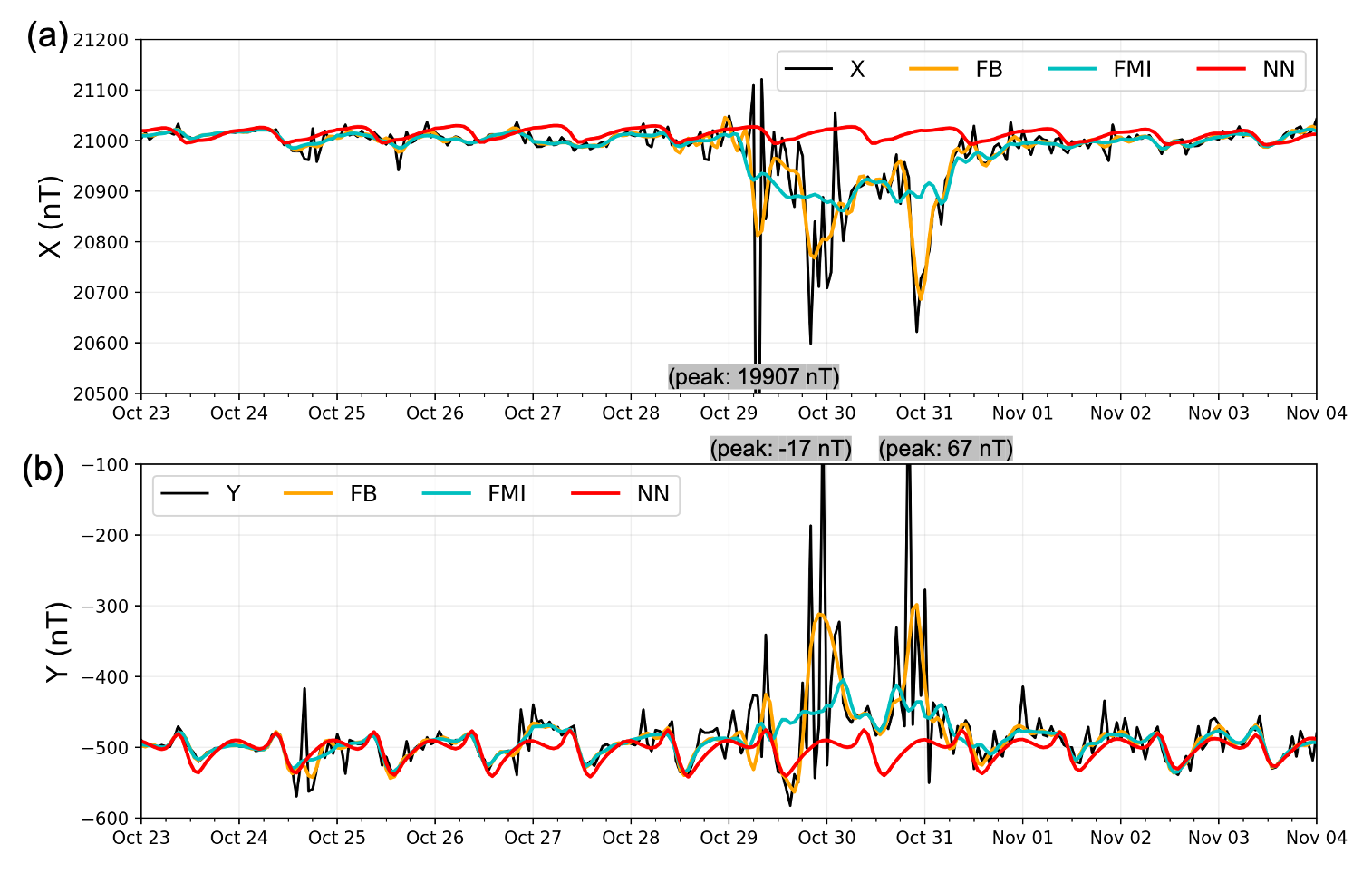}}
 \caption{Comparison between the hourly measurements at CLF (black) and the various geomagnetic baselines: filter (FB; yellow), FMI (cyan), and NN (red) for the Halloween 2003 geomagnetic storm shown for (a) $X$ and (b) $Y$ components.}
 \label{fig:Halloween2003}
\end{figure}

Finally, we compare our approach to the existing methods. Figure~\ref{fig:Halloween2003} shows the hourly measurements (black), the filter baseline (FB, yellow), the FMI baseline (cyan), and our NN baseline (red) produced for the Halloween 2003 event. Figures~\ref{fig:Halloween2003}a and \ref{fig:Halloween2003}b show the $X$- and $Y$-components at CLF, respectively. Here, the NN baseline is produced daily with the lead time of 1 day similar to Figure~\ref{fig:forecast}b. The geomagnetic storm perturbations are clearly visible from October 28 to November 1, with a maximum drop in $X$ of about 1000 nT and a maximum enhancement in $Y$ of about 500 nT compared to the quiet-level variation preceding the event. In the absence of the geomagnetic storm before and after the Halloween event, we find that the FB and FMI baselines follow closely the hourly measurements. The NN baseline, although following the measurements to certain extent, shows some departure or offset from the measurement values, e.g., on October 26 - 27 in Figure~\ref{fig:Halloween2003}a. This offset is likely due to the secular trend projection or the lack of atmospheric contribution in the daily quiet variation prediction. This result suggests that the FB and FMI baselines may better represent the quiet variation than the NN baseline in the absence of solar-origin perturbations.

During the Halloween event from October 28 to November 1, the three baselines differ significantly. The FB and FMI baseline follow the storm variation with the FB being more sensitive to the perturbation compared to the FMI baseline. Regarding the global extrema in $Y$ (at about 67 nT) preceding October 31 in Figure~\ref{fig:Halloween2003}b, for instance, the FB extrema is about 1/3 while the FMI extrema is about 20\% of the $Y$ magnitude. Meanwhile, our NN baseline remains insensitive to the storm variation throughout the event. When characterizing the magnitude of the storm perturbation by subtracting the baseline from the measurement, one would underestimate the actual magnitude if the FB or the FMI baseline is used. In contrast, a better estimate can be obtained using our approach. This result shows that our baseline can robustly characterize the quiet variations during storm perturbation as long as we assume that quiet variations are the same even during a storm. 


\section{Summary and perspectives} \label{sec:conclusions}

We propose a new approach for producing a geomagnetic baseline for robust characterization of the quiet, time-dependent variation in the absence of solar-driven perturbations. As a first step, we demonstrate our approach using ground magnetic data from CLF. The method consists in the neural network prediction of the daily quiet variation, mostly driven by the solar-quiet ionospheric current at mid-latitude, and the extrapolation of the secular variation, owing to the internal geomagnetic field change, using linear regression based on the past 90-day variation. This geomagnetic baseline essentially represents the quiet variations that vary with the (local) intrinsic geomagnetic variability, solar cycles, seasons, and daily variability.

Using the correlation tools, we deduce key parameters consisting in geometrical parameters (LT, SZA, \textit{DistSE}) specific to CLF and the solar irradiance that dominantly correlate to the daily variation. Using the SHAP values, we find similar conclusions for the daily variation during a quiet period. Furthermore, this latter method clearly shows that the perturbed (storm) variation is also correlated to the solar wind and IMF conditions as expected. These approaches help us to identify key parameters that correlate to the daily quiet variation. Physically, the daily quiet variation is driven by the Sq current on the sunlit-side of the ionosphere over mid-latitude regions. Since our neural network yields qualitatively desirable results, it implies that our input variables can be use as proxies indicative of the Sq current variation.

Using the choice of parameters above as inputs and the daily filter as outputs, we develop a LSTM neural network that can robustly predict the daily quiet variation regardless of the solar activities. The neural network is trained using at least 11 years of 1-hour cadence data and updated using the walk forward approach. Our neural network can yield prediction of the daily quiet variation 1 hour in advance based on the past 12-hour history. Furthermore, using time-shifted $F_{10.7}$, our prediction can be obtained 1 day and 27 days in advance. Combining with the linear extrapolation of the secular trend, our method yields a new geomagnetic baseline that is convenient for operational implementation. As demonstrated for the Halloween 2003 event, our baseline is insensitive to external perturbations unlike the FB and FMI baselines. Nevertheless, our baseline shows some offset from the measurements during the quiet time, likely due to the absence of the quiet atmospheric sources and errors in the secular trend extrapolation. Future improvements should include such quiet sources to better represent the daily quiet variation in the absence of solar-origin perturbations. Also, the secular trend prediction can be improved.

In conclusions, we present a new neural network-based approach for automatic generation of a new geomagnetic baseline prediction that is robust against solar-origin perturbations. Our approach is convenient to implement and is scalable to other magnetic observatories. It thus offers an alternative approach for development of a new generation of geomagnetic indices with fine scale and high time resolution for operational space weather. 

\section{Open Research}

The CLF magnetic observatory data are available from Bureau Central de Magnétisme Terrestre data repository (\url{http://doi.org/10.17616/R31NJMXR}) \cite{bcmt} and at Intermagnet data repository (\url{http://doi.org/10.17616/R3XK82}). We acknowledge use of NASA/GSFC’s Space Physics Data Facilities (\url{http://doi.org/10.17616/R3P301}): OMNIWeb (\url{http://doi.org/10.17616/R3TH0D}), CDAWeb (\url{http://doi.org/10.17616/R39H0R}) and OMNI data. Examples of the Python code and data used for producing the results in this work can be retrieved from \url{https://doi.org/10.5281/zenodo.13881560}.

\section{Acknowledgements}
The authors acknowledge the support of the AID (Agence de l’Innovation et de la Defense) and the French ANR (Agence Nationale de la Recherche), under grant ANR-20-ASTC-0030 (project ANR ASTRID PRISMS). Work at IRAP is supported by Centre National de la Recherche Scientifique (CNRS) through the Programme National Soleil-Terre of the Institut National Terre et Univers (PNST / INSU), Centre National d’Etudes Spatiales (CNES) and the University of Toulouse III (UT3). Work at EOST is supported by CNRS, CNES and the University of Strasbourg. RK acknowledges the postdoctoral fellowship from CNES.

\section*{Appendix: Brief description of LSTM neural networks}

\begin{figure}[ht]
\centerline{\includegraphics[width=0.6\textwidth]{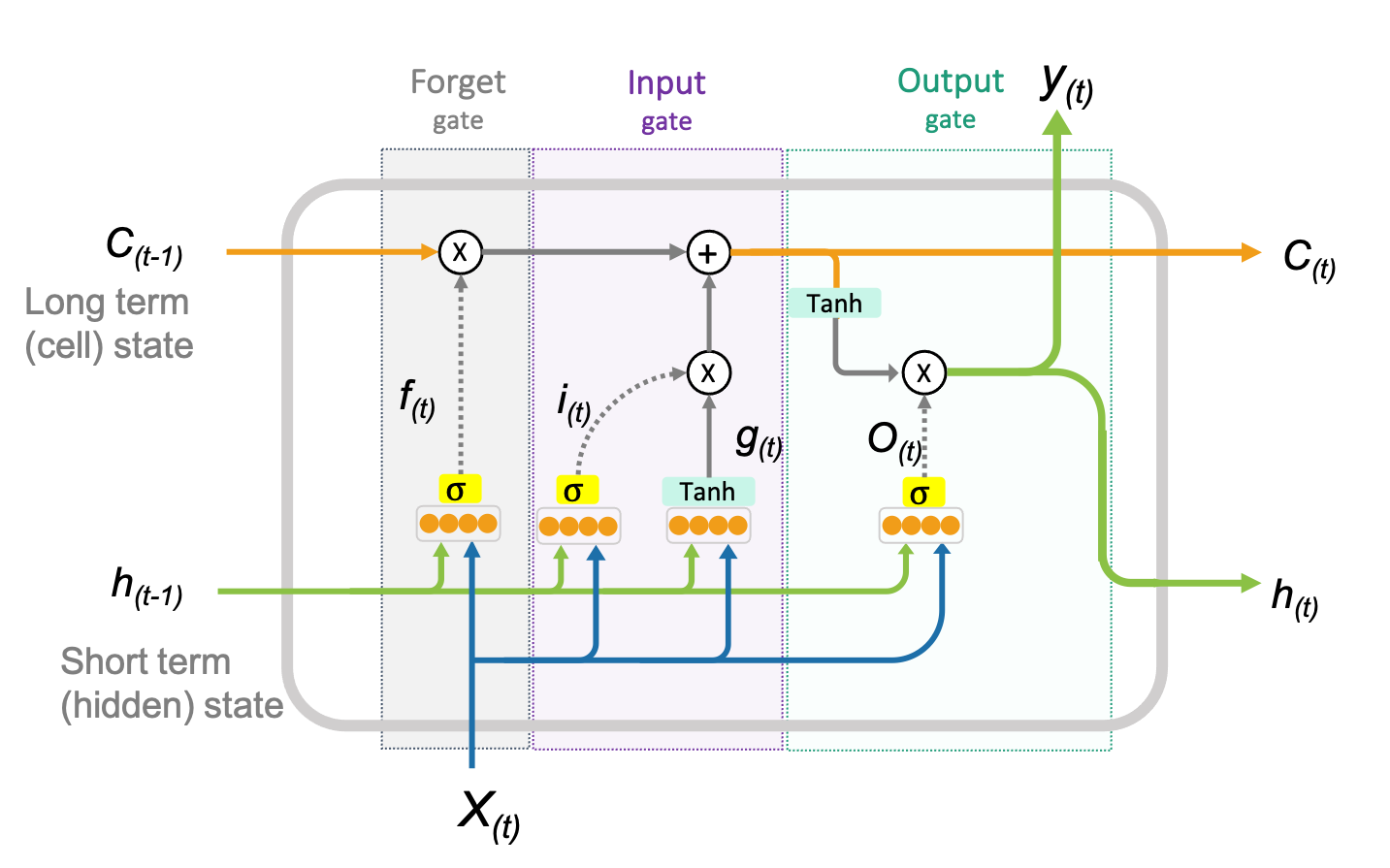}}
{Figure A1. Schematic of the internal structure of a LSTM unit. $\sigma$ represents a layer of neurons with sigmoid activations, Tanh represents a layer of neurons with tanh activations, $+$ represents vector addition, and $\times$ represents vector multiplication.}
\label{fig:LSTM-cell}
\end{figure}

The core of a LSTM unit is a cell where the short-term memory (called the activation) is stored and propagated forward. Figure~\ref{fig:LSTM-cell} shows the internal structure of a LSTM cell. The propagation of the activation vectors within the cell is controlled by the forget gate, the input gate, and the output gate. For each time step ($t$), the parameters within the cell are calculated as follows.

\begin{eqnarray*}
f_{(t)} &=& \sigma(W_{xf}^T X_{(t)} + W_{hf}^T h_{(t-1)} + b_f), \\
i_{(t)} &=& \sigma(W_{xi}^T X_{(t)} + W_{hi}^T h_{(t-1)} + b_i), \\
g_{(t)} &=& \tanh(W_{xg}^T X_{(t)} + W_{hg}^T h_{(t-1)} + b_g), \\
o_{(t)} &=& \sigma(W_{xo}^T X_{(t)} + W_{ho}^T h_{(t-1)} + b_o), \\
c_{(t)} &=& f_{(t)} \otimes c_{(t-1)} + i_{(t)} \otimes g_{(t)},\\
y_{(t)} &=& h_{(t)} = o_{(t)} \otimes \tanh(c_{(t)}),
\end{eqnarray*}

where $X_{(t)}$ is the input vector, $f_{(t)}$ is the forget gate's activation vector, $i_{(t)}$ is the input gate's activation vector, $g_{(t)}$ is the current entry vector, $h_{(t)}, y_{(t)}$ is the hidden state or output vector, $c_{(t)}$ is the cell state vector, $\otimes$ represents the element-wise multiplication, $\sigma$ is the sigmoid function (also known as logistic), $W_k$ where $k \in (f, i, g, o)$ are weights matrices, $b_k$ are bias vectors, and $\tanh$ is the hyperbolic tangent function. Indeed, a LSTM unit is a network in itself as it contains multiple layers of neurons as depicted in Figure~\ref{fig:LSTM-cell}.

\section*{Supplementary Information}

Here we provide information in complementary to Section 4 of the paper regarding choices of the neural network architecture. We set up several experiments as shown in Table S1 in terms of number of nodes in the first LSTM (hidden) layer (also called the input layer) number of nodes in the last LSTM layer, number of additional hidden LSTM layers, and the dropout layer. The neural networks were trained with 1h cadence data from 1997 to 2007, validated using data in 2008, and tested using data in 2009 (e.g., the solar minimum). The sequential input length was set to 12 hours (see more below). The loss, taken as mean square error (MSE), was computed from the original daily filter and the prediction from the neural network. Due to the stochastic nature of the optimization during the neural network training, the MSE improvement when increasing numbers of nodes or layers is not necessarily linear. The best architecture is highlighted in bold. Its schematic illustration is shown in Figure 4c of the main paper. We note that the choice of parameters tested here are not exhaustive. A better MSE may be found using different architectures and hyperparameters from those shown in the main paper. Nevertheless, the model is chosen provides rather satisfactory results as shown in Section 5.

Additionally, we also perform experiments for the length of the sequential inputs as shown in Table S2 based on the best model obtained in Table S1. 6, 8, 12, 16, and 24 hours of inputs were tested. The best MSE is found using 12 hours. 

\noindent\textbf{Table S1.} Experiments with different architectures of the neural network. Various numbers of nodes in the first LSTM (hidden) layer and last LSTM layer, along with the numbers of hidden LSTM layers were tested. Different dropout ratios in the dropout layer (if used) were also tested. MSE and CPU time (i.e., time spent for the training) are provided. The best model with the minimum MSE is highlighted in bold.

\begin{table}[ht]
\centering
\begin{tabular}{|p{0.13\textwidth}|p{0.13\textwidth}|p{0.13\textwidth}|p{0.13\textwidth}|p{0.13\textwidth}|p{0.13\textwidth}|}
\hline
{Number of nodes in first layer} & {Number of nodes in last layer} & {Number of hidden layers (numbers of nodes)} & {Dropout layer (dropout ratio)} & {MSE (nT$^2$)} & {CPU time}   \\ \hline
16  & 8  & - & -   & 4.73E-04  & 3min 28s   \\
32  & 8  & - & -   & 4.18E-04  & 6min 48s   \\
32  & 8  & 1 (8)          & -   & 4.28E-04  &6min 53s   \\
64  & 8  & - & -   & 4.48E-04 & 7min 14s    \\
64  & 32 & - & -   & 4.34E-04  & 8min 19s  \\
128 & 64 & 1 (64)          & -   & 4.55E-04 &14min 46s    \\
128 & 64 & 2 (64)          & -   & 4.46E-04  & 14min 3s   \\
100 & 50 & - & -   & 4.19E-04   &12min 44s  \\
100 & 50 & - & 0.2 & 4.38E-04   & 11min 1s  \\
100 & 50 & 1 (50)          & -   & 4.67E-04  &5min 28s   \\
100 & 50 & 1 (50)          & 0.2 & 4.17E-04  &11min 50s   \\
100 & 50 & 2 (50)          & -   & 4.68E-04  &  7min 3s \\
\textbf{100} & \textbf{50} & \textbf{2 (50)}          & \textbf{0.2} & \textbf{4.01E-04}   & \textbf{25min 59s}  \\
100 & 50 & 2 (50)          & 0.3 & 4.05E-04 & 31min 59s    \\
100 & 50 & 2 (50)          & 0.4 & 4.27E-04  &16min 14s  \\
100 & 50 & 3 (50)          & -   & 4.64E-04  & 8min 3s   \\
100 & 50 & 3 (50)          & 0.2 & 4.43E-04  & 17min 20s  \\ \hline
\end{tabular}
\end{table}

\noindent\textbf{Table S2.} Experiments using different lengths of the input sequences for the neural network. MSE and CPU time (i.e., time spent for the neural network training) are provided. The best MSE is found using 12 hours as highlighted in bold.

\begin{table}[ht]
\centering
\begin{tabular}{|p{0.22\textwidth}|p{0.22\textwidth}|p{0.22\textwidth}|}
\hline 
{Sequential input lengths} & {MSE (nT$^2$)} & {CPU time}      \\ \hline
 6h       & 4.74E-04 &  5min 31s \\
 8h       & 4.26E-04 & 19min 16s  \\
 \textbf{12h}      & \textbf{4.01E-04} & \textbf{25min 59s}   \\
 16h      & 4.03E-04 & 27min 19s \\
 24h      & 4.17E-04 &  31min 15s \\ \hline
\end{tabular}
\end{table}

\bibliographystyle{unsrtnat}
\bibliography{template}  






\end{document}